\documentclass[%
 aip, 
 amsmath, amssymb, 
 reprint, %
]{revtex4-1}

\usepackage{graphicx}
\usepackage{dcolumn}
\usepackage{bm}
\usepackage[utf8]{inputenc}
\usepackage[T1]{fontenc}
\usepackage{mathptmx}
\usepackage{etoolbox}
\usepackage{caption}
\makeatletter
\def\@email#1#2{%
 \endgroup
 \patchcmd{\titleblock@produce}
  {\frontmatter@RRAPformat}
  {\frontmatter@RRAPformat{\produce@RRAP{*#1\href{mailto:#2}{#2}}}\frontmatter@RRAPformat}
  {}{}
}%
\makeatother
\begin{document}

\preprint{AIP/123-QED}

\title{Moving mesh FSI approach for VIV simulation based on DG method with AMR technique}
\affiliation{College of Shipbuilding Engineering, Harbin Engineering University, Harbin 150001, China}

\author{Jia-Jun Zou}
\altaffiliation{} 

\author{Yun-Long Liu}
\email{yunlong\_liu@hrbeu. edu. cn}
\altaffiliation{} 

\author{Qi Kong}
\altaffiliation{} 

\author{A-Man Zhang}
\altaffiliation{} 

\date{\today}

\begin{abstract}
  Vortex-induced vibration (VIV) remains a fundamental yet computationally
   challenging problem in computational fluid dynamics (CFD). This study develops a moving mesh Fluid-structure interaction (FSI) approach within a Runge-Kutta 
    Discontinuous Galerkin (RKDG) adaptive mesh refinement (AMR) framework. The viscous term in the compressible Navier-Stokes (NS) equations is discretized using 
    the high-order Interior Penalty Discontinuous Galerkin (IPDG) 
    method. In addition to the above, key numerical advancements encompass 
    the rigorous derivation of the Lax-Friedrichs (L-F) numerical flux formulation 
    tailored for moving meshes, an enhanced AMR-driven nodal correction 
    methodology designed for curved surface geometries, 
    and the implementation of a ghost-node boundary condition treatment scheme 
    to address dynamic mesh motion. Numerical validation proceeds through three phases: 
    First, Couette flow simulations confirm the IPDG method's spatial 
    convergence order. Subsequent analysis of unsteady flow past a cylinder 
    demonstrate the AMR framework's efficacy in resolving vortex-dominated flow. 
    Finally, six VIV benchmark cases are simulated using third-order IPDG discretization, 
    establishing the proposed FSI approach's accuracy. Furthermore, synthetic jets (SJs) flow control is investigated through 
    four frequency-variant SJs configurations. The results reveal that SJs can achieve completely 
    VIV suppression at a low actuation frequency, while higher actuation 
    frequencies reduce suppression efficiency due 
    to the energy of the SJs is more in the form of acoustic wave. 
\end{abstract}

\maketitle

\section{INTRODUCTION}

FSI phenomena are prevalent across diverse engineering disciplines, observed in 
physical systems ranging from submarine pipeline oscillations under hydrodynamic forces \cite{xu2022multispan, zhang2022numerical, lu2025vortex} 
to aeroelastic instabilities in aircraft structures \cite{suleman2006active, henshaw2007non, chen2025investigation}
and wind-induced vibrations in long-span bridge systems \cite{chen2003evolution, cai2004wind, zhang2021wind}. VIV serves as a paradigmatic FSI phenomenon, where transverse VIV 
has become a benchmark \cite{AHN2006671, borazjani2008curvilinear} for validating FSI algorithms due to its geometrically 
simple configuration and rich phenomenological manifestations. This canonical problem 
is commonly investigated using an elastically mounted rigid cylinder model.

Extensive research has historically focused on advancing numerical simulations of VIV. 
Ahn and Kallinderis \cite{ahn2006strongly} employed the Arbitrary Lagrangian-Eulerian Finite Volume Method (ALE-FVM) 
coupled with an artificial compressibility approach in 2006 for VIV analysis. Subsequent developments include Xie et al. \cite{xie2012numerical} who investigated flexible cylindrical 
structures using FVM in 2012, and Han et al. \cite{jiao2021vortex} who implemented the Immersed Boundary-Lattice 
Boltzmann Method (IB-LBM) in 2021 for VIV simulations. Howe ver, existing research predominantly remains confined 
to second-order accurate numerical schemes in FSI simulations. The potential of high-order accuracy methods, 
which exhibit superior numerical characteristics with reduced dispersion and dissipation, 
remains underexplored in FSI simulations. These high-order accuracy methods enable finer resolution of flow field 
structures under same grid compared to conventional second-order accuracy approaches such as second-order FVM. 
Building upon these foundations, this study proposes a high-order accuracy FSI approach 
based on the RKDG \cite{cockburn2001runge, liu2016positivity, zhu2013runge} AMR framework. 

The DG method, a prominent high-order finite element approach, features a compact scheme that facilitates AMR and parallel computing.
When applying DG methods to NS equations, researchers have developed distinct discretization approaches 
for viscous terms, including: the Bassi-Rebay 1 \cite{bassi1997high, manzanero2018bassi} (BR1), 
Bassi-Rebay 2 \cite{bosnyakov2019unsteady, arnold2002unified} (BR2), 
IPDG \cite{arnold1982interior, hartmann2008optimal, heimann2013unfitted}, Local DG
\cite{kaltenbach2023local, baccouch2012local, cockburn1998local}  (LDG), Compact DG 
\cite{peraire2008compact, brdar2012compact} (CDG), and Direct DG 
\cite{cheng2016direct, zhang2019direct, liu2009direct} (DDG). 
Among these variants, the IPDG method demonstrates superior computational efficiency and implementation simplicity. 
Consequently, this study adopts the IPDG method for discretizing the viscous terms 
of NS equations. 

Although the DG method exhibits significant advantages in high-order accuracy, 
its computational efficiency remains constrained by considerable computational costs. It is noteworthy that the compact stencil characteristic of DG schemes makes them 
particularly suitable for integration with h-adaptive mesh refinement techniques 
\cite{schaal2015astrophysical, papoutsakis2018efficient, wang2009adjoint, liu2025unstructuredblockbasedadaptivemesh}. This approach selectively refines or coarsens grid elements based on characteristic physical quantities within cells, 
such as velocity divergence or vorticity magnitude. By dynamically adapting the mesh resolution, 
this strategy achieves an optimal balance between computational accuracy and efficiency, 
enabling precise capture of critical flow features like shock waves and vortex cores while substantially 
reducing computational resource requirements. 

Within the algorithmic theoretical framework, 
this study methodically constructs a computational architecture for solving compressible NS 
equations on moving mesh (Sec.\ref{sec2.1}). The theoretical development commences with spatial discretization 
employing the high-order IPDG method under the RKDG AMR framework (Sec.\ref{sec2.2}). Subsequently, a rigorous mathematical derivation of the 
L-F numerical flux \cite{toro2013riemann} formulation specifically adapted for moving mesh systems is presented (Sec.\ref{sec2.3}). An enhanced 
AMR-driven nodal correction methodology is further proposed to address curved boundary geometries 
with increased precision (Sec.\ref{sec2.4}). The implementation incorporates a ghost-node methodology for moving boundary 
condition enforcement in moving mesh configurations (Sec.\ref{sec2.5}). The FSI computational protocol 
for VIV is systematically formulated (Sec.\ref{sec2.6}), incorporating a third-order Total Variation Diminishing Runge-Kutta (TVD-RK) temporal integration 
scheme to ensure enhanced numerical stability during coupled system evolution. 

The numerical validation section follows a systematic verification protocol. Initially, the IPDG method is 
implemented on fixed grids to solve the NS equations, 
with the classical Couette flow benchmark rigorously validating the high precision and computational 
reliability of the developed solver (Sec.\ref{coue}). Subsequent analysis of unsteady flow past a cylinder demonstrate 
the AMR framework's efficacy in resolving vortex-dominated flow (Sec.\ref{cy flow}). To further evaluate the approach's 
engineering applicability, six VIV benchmark cases are 
simulated using third-order IPDG discretization. Quantitative comparisons with authoritative literature data confirm the excellent computational accuracy 
and engineering robustness of the proposed moving mesh FSI approach. Ultimately, 
SJs flow control \cite{glezer2002synthetic} is investigated through four frequency-variant SJs configurations, 
which not only substantiate the jet control strategy as a stable and efficient solution for VIV suppression, 
but more importantly unravel the intrinsic correlation between jet actuation frequency and vibration mitigation 
effectiveness (Sec.\ref{SJs-VIV}). 
\section{NUMERICAL METHODS AND TECHNIQUES}
\subsection{\label{sec2.1}Fluid dynamics equations in a moving mesh}
In the present VIV simulation, 
two distinct reference frames are employed: 
a stationary earth- fixed reference frame and 
a moving reference frame that moves synchronously with 
the rigid cylindrical structure.
Within the RKDG AMR framework, this study develops a moving mesh FSI approach. 
Built upon the compressible NS equations, the work theoretically derives generalized hydrodynamic 
equations in the Eulerian reference frame that govern fluid motion within dynamically moving mesh. 
A distinguishing feature of these governing equations is the incorporation of an additional transport term 
originating from moving mesh, which constitutes a theoretical extension beyond the classical compressible 
NS equations and substantially enhances the approach's capability in modeling dynamic boundary problems. 

In the Cartesian coordinate system, the two-dimensional compressible Navier-Stokes equations neglecting body forces and volumetric heat sources are presented as follows:
\begin{equation}
  \frac{\partial \mathrm{\mathbf{U}}}{\partial t} +\nabla \cdot (\mathrm{\mathbf{F}}^c(\mathrm{\mathbf{U}})-\mathrm{\mathbf{F}}^v(\mathrm{\mathbf{U}}, \nabla \mathrm{\mathbf{U}}))=0, 
\label{eq:1}
\end{equation}
where $\mathrm{\mathbf{U}}$ denotes the conserved variables, $\mathrm{\mathbf{F}}^c=(\mathrm{\mathbf{F}}^c_x, \mathrm{\mathbf{F}}^c_y)$ represents the convective flux, and 
$\mathrm{\mathbf{F}}^v=(\mathrm{\mathbf{F}}^v_x, \mathrm{\mathbf{F}}^v_y)$ corresponds to the viscous flux. Their specific forms are defined as follows:

\begin{eqnarray}
  \mathrm{\mathbf{U}}= \left[
  \begin{array}{c}
    \rho \\
    \rho u\\
    \rho v\\
    \rho E
  \end{array}
  \right], 
  \mathrm{\mathbf{F}}^c_x = \left[
    \begin{array}{c}
      \rho u\\
      \rho u^2+p\\
      \rho uv\\
      u(\rho E+p)
    \end{array}
    \right], 
    \mathrm{\mathbf{F}}^c_y = \left[
    \begin{array}{c}
      \rho v\\
      \rho uv\\
      \rho v^2+p\\
      v(\rho E+p)
    \end{array}
    \right], \nonumber\\ 
    \mathrm{\mathbf{F}}^v_x = \left[
      \begin{array}{c}
        0\\
        \tau_{xx}\\
        \tau_{xy}\\
        u\tau_{xx}+v\tau_{xy}+k_c\frac{\partial T}{\partial x} 
      \end{array}
      \right], 
      \mathrm{\mathbf{F}}^v_y = \left[
        \begin{array}{c}
          0\\
          \tau_{yx}\\
          \tau_{yy}\\
          u\tau_{yx}+v\tau_{yy}+k_c\frac{\partial T}{\partial y} 
        \end{array}
        \right], 
        \label{eq:2}
\end{eqnarray}
where $\rho$, $\mathrm{\mathbf{V}}=(u, v)$, $p$, $E$, and $T$ denote the density, the velocity vector, the
pressure, the specific total energy, and the temperature, respectively.
Moreover, $k_c$ is the thermal conductivity coefficient. For the Newtonian fluid, 
the viscous stress tensor is given by:
\begin{equation}
  \mathrm{\mathbf{\tau}}=\mu\left[
     \nabla \mathrm{\mathbf{V}}+
     (\nabla \mathrm{\mathbf{V}})^{\mathrm{T}}-\frac{2}{3}(\nabla \cdot \mathrm{\mathbf{V}})\mathrm{\mathbf{I}}
    \right], 
    \label{eq:3}
\end{equation}
where $\mathrm{\mathbf{I}}$ is the identity matrix, $\mu$ is the dynamic viscous coefficient.
 Since the specific internal energy 
$e$ (per unit mass) and temperature $T$ satisfy $e = C_v T$, we have:
\begin{equation}
  k_cT = \frac{\mu\gamma}{\mathrm{Pr}}e, 
  \label{eq:4}  
\end{equation}
where $\gamma$ is the ratio of the
specific heats, $\mathrm{Pr}$ is the Prandtl number of the fluid (for air, the laminar Prandtl number is taken as 0.72). The total energy per unit mass $E$ of the fluid can be expressed as:
\begin{equation}
  E = \frac{|\mathrm{\mathbf{V}}|^2}{2}+e. 
  \label{eq:5}  
\end{equation}

For compressible fluids, the Tammann equation of state \cite{ivings1998riemann} (eos) is 
introduced to describe the relationship between the internal energy and pressure of the fluid:
\begin{equation}
  p = \rho e({\gamma-1})-\gamma P_w, 
  \label{eq:6}  
\end{equation}
where $P_w$ is the reference pressure of the fluid. In the test cases,
the fluid is assumed to be an ideal gas, i. e. , $\gamma=1. 4$ and $P_w=0$ . 

Ahn and Kallinderis \cite{ahn2006strongly} (2006) implemented VIV simulations 
using a moving reference frame in their ALE-FVM approach, 
where the coordinate system was attached to moving boundaries. In contrast, our methodology employs a stationary earth-fixed reference frame. 
The computational grid moves synchronously with the rigid cylindrical structure in translational motion,
 maintaining velocity consistency 
$\mathrm{\mathbf{V}}_s= \mathrm{\mathbf{V}}_{mesh}=(u_{mesh}, v_{mesh})$. 
Based on Eq.~(\ref{eq:1}), 
the modified governing equations for fluid dynamics in this translational moving mesh system can be expressed as:
\begin{equation}
  \frac{\partial^* \mathrm{\mathbf{U}}}{\partial^* t} +
  \nabla \cdot (\mathrm{\mathbf{F}}^c(\mathrm{\mathbf{U}})-
  \mathrm{\mathbf{F}}^v(\mathrm{\mathbf{U}}, \nabla \mathrm{\mathbf{U}}))=\mathrm{\mathbf{V}}_{mesh} \cdot \nabla \mathrm{\mathbf{U}}, 
\label{eq:7}
\end{equation}
The fundamental relationship between 
$\frac{\partial^* \mathrm{\mathbf{U}}}{\partial^* t}$ and $\frac{\partial \mathrm{\mathbf{U}}}{\partial t}$ defined as 
$\frac{\partial^* \mathrm{\mathbf{U}}}{\partial^* t}= \frac{\partial \mathrm{\mathbf{U}}}{\partial t}
+\mathrm{\mathbf{V}}_{mesh} \cdot \nabla \mathrm{\mathbf{U}}$
. The term $\mathrm{\mathbf{V}}_{mesh} \cdot \nabla \mathrm{\mathbf{U}}$ on the right-hand side physically represents the additional transport term induced by the mesh motion. Consequently, Eq.~(\ref{eq:7}) can be simplified to:
\begin{equation}
  \frac{\partial^* \mathrm{\mathbf{U}}}{\partial^* t} +
  \nabla \cdot (\mathrm{\mathbf{F}}^{\tilde{c}}(\mathrm{\mathbf{U}})-
  \mathrm{\mathbf{F}}^v(\mathrm{\mathbf{U}}, \nabla \mathrm{\mathbf{U}}))=-\mathrm{\mathbf{U}} \nabla \cdot \mathrm{\mathbf{V}}_{mesh}, 
\label{eq:8}
\end{equation}
the new convective flux term is formally defined as:
\begin{equation}
\begin{aligned}
  \mathrm{\mathbf{F}}^{\tilde{c}}(\mathrm{\mathbf{U}})&=(\mathrm{\mathbf{F}}^{\tilde{c}}_x(\mathrm{\mathbf{U}}), 
  \mathrm{\mathbf{F}}^{\tilde{c}}_y(\mathrm{\mathbf{U}}))\\ 
   &=(\mathrm{\mathbf{F}}^c_x(\mathrm{\mathbf{U}})-u_{mesh}\mathrm{\mathbf{U}}, 
   \mathrm{\mathbf{F}}^c_y(\mathrm{\mathbf{U}})-v_{mesh}\mathrm{\mathbf{U}}). 
\end{aligned} 
\label{eq:9}
\end{equation}

\subsection{\label{sec2.2}Review of the RKDG method}
The RKDG method refers to a numerical discretization approach where 
temporal discretization is implemented using the TVD-Runge-Kutta scheme 
, while spatial discretization employs the DG method, 
thereby achieving spatiotemporal discretization of partial differential equations. Due to the high-order accuracy and inherent stability characteristics of the RKDG framework, 
we select this computational paradigm to solve Eq.~(\ref{eq:8}). 

Following the fundamental principles of the DG finite element framework, 
we seek the projection $\mathrm{\mathbf{U}}_h$ of the true unknown solution $\mathrm{\mathbf{U}}$ 
within the discontinuous finite element space over the computational domain. 
Let $\Omega$ denote the fluid domain with boundary $\partial \Omega$. 
The domain $\Omega$ is discretized into non-overlapping elements $\Omega_e$, 
where each element boundary $\partial \Omega_e$ is associated with a unit outward 
normal vector $\mathrm{\mathbf{n}}=(n_x, n_y)$. Following the aforementioned framework, 
we formally introduce the DG finite element space $V_h$, defined as:

\begin{equation}
V_h = \left\{ {\varphi _h} \in L^2( \Omega ):\varphi _h|_{\Omega _e}
\in {P}^s(\Omega _e), \forall \Omega _e \in \Omega  \right\}, 
\label{eq:10}
\end{equation}
where ${P}^s(\Omega _e)$ denotes the space of polynomial functions of degree $s$ on each element $\Omega_e$. 
Within the DG framework, we first construct a set of 
basis functions $\{\varphi_i\}_{i=1}^N$ defined on the polynomial space ${P}^s(\Omega_e)$, where the dimension $N$ is determined by $N = \frac{1}{2}(s+1)(s+2)$. 
Following the DG discretization strategy, the numerical solution $\mathrm{\mathbf{U}}_h$ within each computational element is expressed as a linear combination of these basis functions:
$\mathrm{\mathbf{U}}_h(\mathbf{x}, t) = \sum_{i=1}^N \mathbf{k}_i(t) \varphi_i(\mathbf{x})$. 
This representation achieves spatiotemporal decoupling of the approximate solution, 
where the time-dependent coefficient vectors $\mathbf{k}_i(t)$ exclusively serve 
as the degrees of freedom (DOFs) within the local element. 
By applying the Galerkin integration procedure to Eq.~(\ref{eq:8}) over an individual element:
\begin{equation}
\begin{aligned}
  & \int\limits_{\Omega_e}{\frac{\partial^* \mathrm{\mathbf{U}}}{\partial^* t}\varphi_i \, dS} \\
 & = -\int\limits_{\Omega_e}{\left[ \nabla \cdot \left( \mathbf{F}^{\!\tilde{c}}(\mathrm{\mathbf{U}}) - \mathbf{F}^{\!v}(\mathrm{\mathbf{U}}, \nabla \mathrm{\mathbf{U}}) \right)\varphi_i - \mathrm{\mathbf{U}} \nabla \cdot \mathbf{V}_{mesh}\varphi_i \right] dS} \\
 & = \int\limits_{\Omega_e}{\left[ \left( \mathbf{F}^{\!\tilde{c}}(\mathrm{\mathbf{U}}) - \mathbf{F}^{\!v}(\mathrm{\mathbf{U}}, \nabla \mathrm{\mathbf{U}}) \right) \cdot \nabla \varphi_i - \mathrm{\mathbf{U}} \nabla \cdot \mathbf{V}_{mesh}\varphi_i \right] dS} \\
 & \quad -\int\limits_{\partial \Omega_e}{\left[ \left( \mathbf{\hat{F}}^{\!\tilde{c}}(\mathrm{\mathbf{U}}) - \mathbf{\hat{F}}^{\!v}(\mathrm{\mathbf{U}}, \nabla \mathrm{\mathbf{U}}) \right) \right] \cdot \mathbf{n} \varphi_i \, dl}, 
\end{aligned}
\label{eq:11}
\end{equation}
the terms $\mathbf{\hat{F}}^{\!\tilde{c}}(\mathrm{\mathbf{U}}) \cdot \mathbf{n}$
 and $\mathbf{\hat{F}}^{\!v}(\mathrm{\mathbf{U}}, \nabla \mathrm{\mathbf{U}}) \cdot \mathbf{n}$
  represent the convective numerical flux and viscous numerical flux, 
  respectively, at element boundaries. 

To preserve flux conservation and scheme consistency, the convective numerical flux $\mathbf{\hat{F}}^{\!\tilde{c}}(\mathrm{\mathbf{U}}) \cdot \mathbf{n}$ is 
constructed using an approximate Riemann solver. 
Given that all investigated cases involve low-Mach-number regimes without strong discontinuities, 
numerical experiments demonstrate negligible computational difference to the selection of different 
convective numerical flux schemes. For computational efficiency optimization, the Lax-Friedrichs numerical flux, 
characterized by its lower computational overhead, is ultimately adopted, 
while the viscous numerical flux 
$\mathbf{\hat{F}}^{\!v}(\mathrm{\mathbf{U}}, \nabla \mathrm{\mathbf{U}}) \cdot \mathbf{n}$ 
adopts the IPDG numerical flux. 
The detailed implementation procedures for these numerical fluxes are comprehensively 
described in Section~\ref{sec2.3}. 

Substituting the numerical solution $\mathrm{\mathbf{U}}_h(\mathbf{x}, t)$ 
into Eq.~(\ref{eq:11}), and considering all test functions $\{\varphi_i\}_{i=1}^N$, 
each element generates a linear system of the form:
\begin{equation}
  \mathbf{M} \dot{\mathbf{K}} = \mathbf{R}, 
  \label{eq:12} 
\end{equation}
where $\mathbf{M}$ is the $N \times N$ mass matrix, $\dot{\mathbf{K}}$
denotes the time derivative of the coefficient matrix, $\mathbf{R}$ 
is the right-hand side matrix. 

The temporal discretization employs the third-order TVD-Runge-Kutta method \cite{gottlieb2001strong} as detailed in Eq.~(\ref{eq:27}). 
The expression for the time step $\Delta t$ adopted in this paper is given by:
\begin{equation}
  \Delta t = \frac{q}{\frac{\sqrt{{{u}^{2}}+{{v}^{2}}}+c_s}{h}+\frac{\mu }{\rho {{h}^{2}}}}, 
  \label{eq:13} 
\end{equation}
in which $q$ is the Courant-Friedrichs-Lewy number, $h$ is the mesh size, 
$c_s$ denotes the speed of sound. 

The RKDG computational framework has thus been established now. During RKDG simulations, numerical oscillations arise when flow discontinuities 
propagate into computational cell interiors. 
Such oscillations may erroneously render inherently positive-definite physical 
quantities (e.g., density and pressure) negative, 
thereby destabilizing the computational process. To address this challenge, researchers have developed various shock-capturing approaches, 
primarily categorized into artificial viscosity methods, 
reconstruction-based methods, 
and limiter techniques. Artificial viscosity methods \cite{hartmann2006adaptive}, 
being the earliest proposed approach, suffer from significant parameter dependence 
and tend to introduce excessive numerical dissipation. 
Recently, 
Liu et al. developed an entropy-stable artificial viscosity method \cite{liu2024non}. 
Reconstruction-based discontinuity-capturing methods \cite{luo2013reconstructed} 
face implementation challenges in complex geometries and three-dimensional 
problems due to their non-compact stencil. In contrast, 
limiter techniques demonstrate superior flexibility in handling complex 
geometries and three-dimensional problems. Considering the physical discontinuities 
that emerge near jet orifices in VIV suppression simulations, this study employs a 
positivity-preserving limiter for strong shock capture within the RKDG framework, 
as proposed by Liu et al. \cite{liu2016positivity} in 2016, 
combined with an unstructured grid-adapted WENO limiter \cite{zhu2013runge} to 
ensure computational stability and accuracy. 

\subsection{\label{sec2.3}numerical flux formulation for moving mesh}
L-F numerical flux \cite{toro2013riemann} formulation for
moving mesh is given by:
\begin{equation}
\mathbf{\hat{F}}^{\tilde{c}}(\mathrm{\mathbf{U}}^{-}, \mathrm{\mathbf{U}}^{+}) \cdot \mathbf{n}
= \frac{1}{2}((\mathbf{F}^{\tilde{c}}(\mathrm{\mathbf{U}}^{-}) +
\mathbf{F}^{\tilde{c}}(\mathrm{\mathbf{U}}^{+})) \cdot \mathbf{n}-\alpha(\mathrm{\mathbf{U}}^{+}-
\mathrm{\mathbf{U}}^{-})), 
\label{eq:14} 
\end{equation}
where $\alpha$ is the maximum absolute value among all eigenvalues of both Jacobian matrices 
$
\tilde{J}_{+} = \frac{\partial \mathbf{F}^{\tilde{c}}(\mathrm{\mathbf{U}}^{+}) \cdot \mathbf{n}}{\partial \mathrm{\mathbf{U}}^{+}}
$
and 
$
\tilde{J}_{-} = \frac{\partial \mathbf{F}^{\tilde{c}}(\mathrm{\mathbf{U}}^{-}) \cdot \mathbf{n}}{\partial \mathrm{\mathbf{U}}^{-}}
$. It is worth noting that, due to the relationship in Eq.~(\ref{eq:9}), the Jacobian matrices 
$
\tilde{J} = \frac{\partial \mathbf{F}^{\tilde{c}}(\mathrm{\mathbf{U}}) \cdot \mathbf{n}}{\partial \mathrm{\mathbf{U}}}
$
and 
$
J = \frac{\partial \mathbf{F}^{c}(\mathrm{\mathbf{U}}) \cdot \mathbf{n}}{\partial \mathrm{\mathbf{U}}}
$
satisfy the following relationship:
\begin{equation}
  \frac{\partial \mathbf{F}^{\tilde{c}}(\mathrm{\mathbf{U}}) \cdot \mathbf{n}}{\partial \mathrm{\mathbf{U}}}
  =\frac{\partial \mathbf{F}^{c}(\mathrm{\mathbf{U}}) \cdot \mathbf{n}}{\partial \mathrm{\mathbf{U}}}-(\mathbf{V}_{mesh} \cdot \mathbf{n}) \mathbf{I}. 
  \label{eq:15} 
\end{equation}

Furthermore, it can be shown that the Jacobian matrices 
$
\tilde{J}= \frac{\partial \mathbf{F}^{\tilde{c}}(\mathrm{\mathbf{U}}) \cdot \mathbf{n}}{\partial \mathrm{\mathbf{U}}}
$
and 
$
J = \frac{\partial \mathbf{F}^{c}(\mathrm{\mathbf{U}}) \cdot \mathbf{n}}{\partial \mathrm{\mathbf{U}}}
$
share the same right eigenvector matrix, and their diagonal eigenvalue matrices 
$\tilde{\lambda} $
(for 
$
\frac{\partial \mathbf{F}^{\tilde{c}}(\mathrm{\mathbf{U}}) \cdot \mathbf{n}}{\partial \mathrm{\mathbf{U}}}
$
) and 
$\lambda $ (for 
$\frac{\partial \mathbf{F}^{c}(\mathrm{\mathbf{U}}) \cdot \mathbf{n}}{\partial \mathrm{\mathbf{U}}}$
) satisfy the following relationship:
\begin{equation}
  \tilde{\lambda}
  =\lambda-(\mathbf{V}_{mesh} \cdot \mathbf{n}) \mathbf{I}. 
  \label{eq:16} 
\end{equation}

Therefore, the characteristic speed $\alpha$ is derived from the following formula:
$\alpha = \max\left\{ (|\mathbf{V} \cdot \mathbf{n}-\mathbf{V}_{mesh} \cdot \mathbf{n}|+c)^+, 
 (|\mathbf{V} \cdot \mathbf{n}-\mathbf{V}_{mesh} \cdot \mathbf{n}|+c)^-\right\}$

 The IPDG viscous numerical flux \cite{hartmann2006symmetric} enhances the weak continuity of the conservative vector 
 at element interfaces by incorporating a penalty term, which is formulated as follows:
\begin{equation}
  \mathbf{\hat{F}}^{\!v}(\mathrm{\mathbf{U}}, \nabla \mathrm{\mathbf{U}}) \cdot \mathbf{n} =
  \frac{\mathbf{F}^{v, +}+\mathbf{F}^{v, -}}{2} \cdot \mathbf{n}-
  C_{ip} \mu \frac{s^2}{h}(\mathrm{\mathbf{U}}^{-}-
  \mathrm{\mathbf{U}}^{+}), 
  \label{eq:17} 
\end{equation}
in which $C_{ip}$ is the penalty coefficient, 
typically chosen as a positive constant. 
It is worth noting that there is currently no universally optimal method for selecting 
$C_{ip}$. Appropriately increasing $C_{ip}$ can enhance the continuity of the solution
 at element interfaces;however, excessively large values of 
$C_{ip}$ may lead to an ill-conditioned problem. In this work, $C_{ip}=4$. 

\subsection{\label{sec2.4}AMR and an enhanced AMR-driven nodal correction methodology for curved surface geometries}

\begin{figure}
  \includegraphics[width=0.45\textwidth]{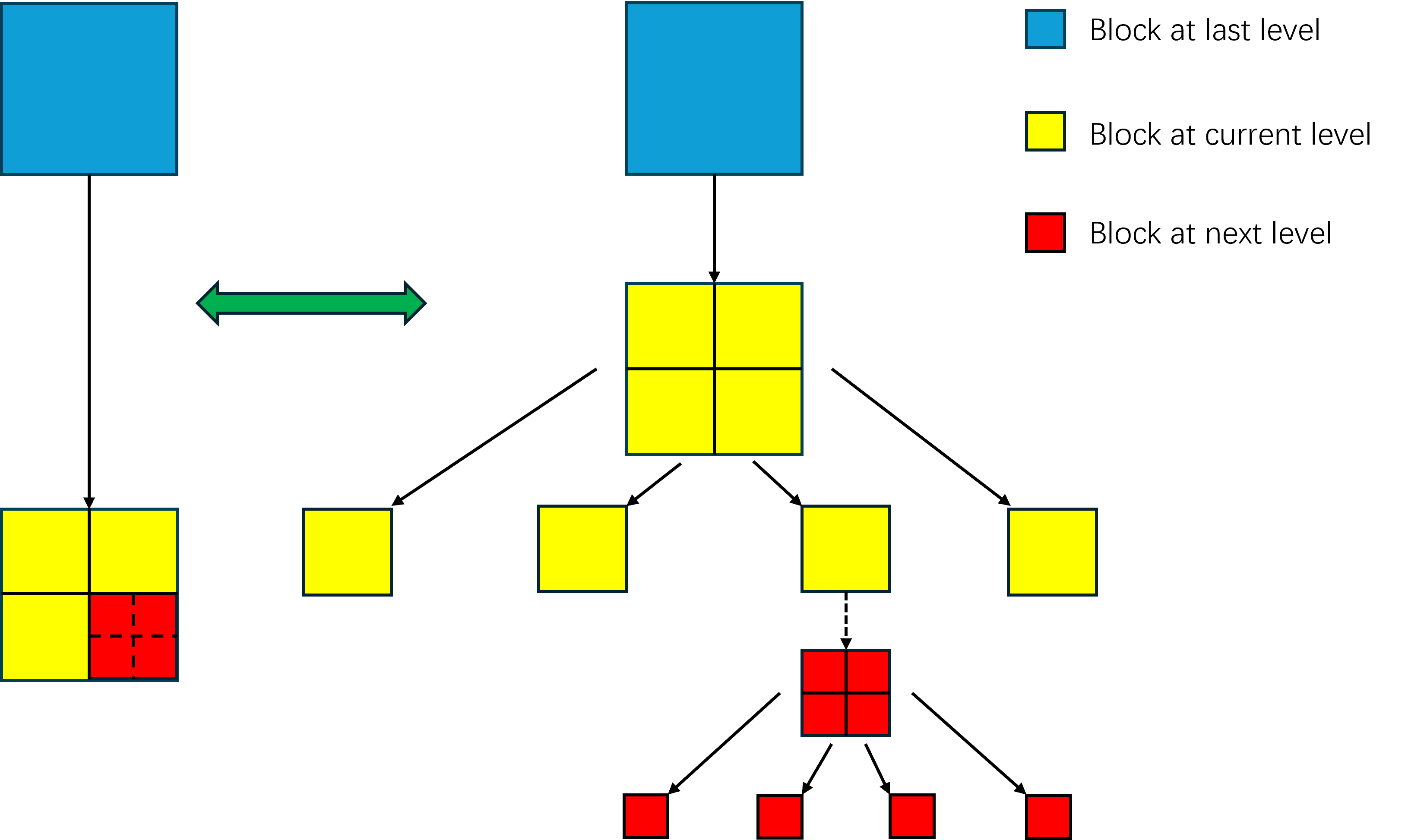}
  \caption{\label{fig2} Schematic of quad-tree AMR architecture. }
\end{figure}

This study employs the FSMESH module within the FSI software FSLAB \cite{2022Liu}
 to implement dynamic adaptive mesh refinement (AMR) under large-scale MPI parallelization. 
 The AMR framework utilizes a block-based quad-tree data structure, 
 where each parent block splits into four child blocks during refinement 
 and conversely merges from four child blocks during coarsening. 
 Each block contains $N_{seg}^2$ computational cells, with explicit DG solutions 
 computed at the individual cell level. 
 Boundary data exchange between blocks is managed through direct 
 communication with neighboring blocks. The AMR configuration specifies 
a user-defined minimum refinement level $A_{min}$ and a user-defined maximum level $A_{max}$. Fig.~\ref{fig2} schematically illustrates this quad-tree AMR architecture. 

\begin{figure}
    \begin{minipage}[b]{0.45\textwidth}
      \centering
      \includegraphics[height=6cm]{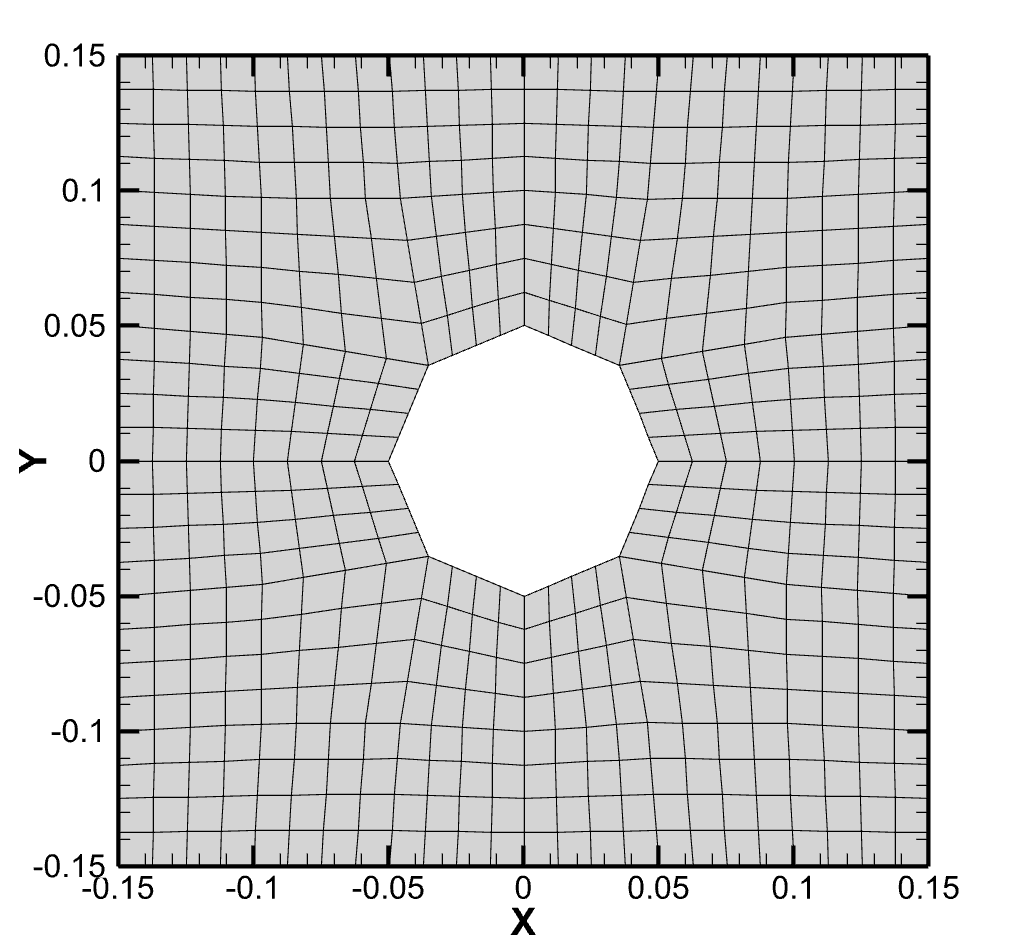} 
      \caption*{(a) without correction}
    \end{minipage}
    \hspace{0.05\textwidth}    
    \begin{minipage}[b]{0.45\textwidth}
      \centering
      \includegraphics[height=6cm]{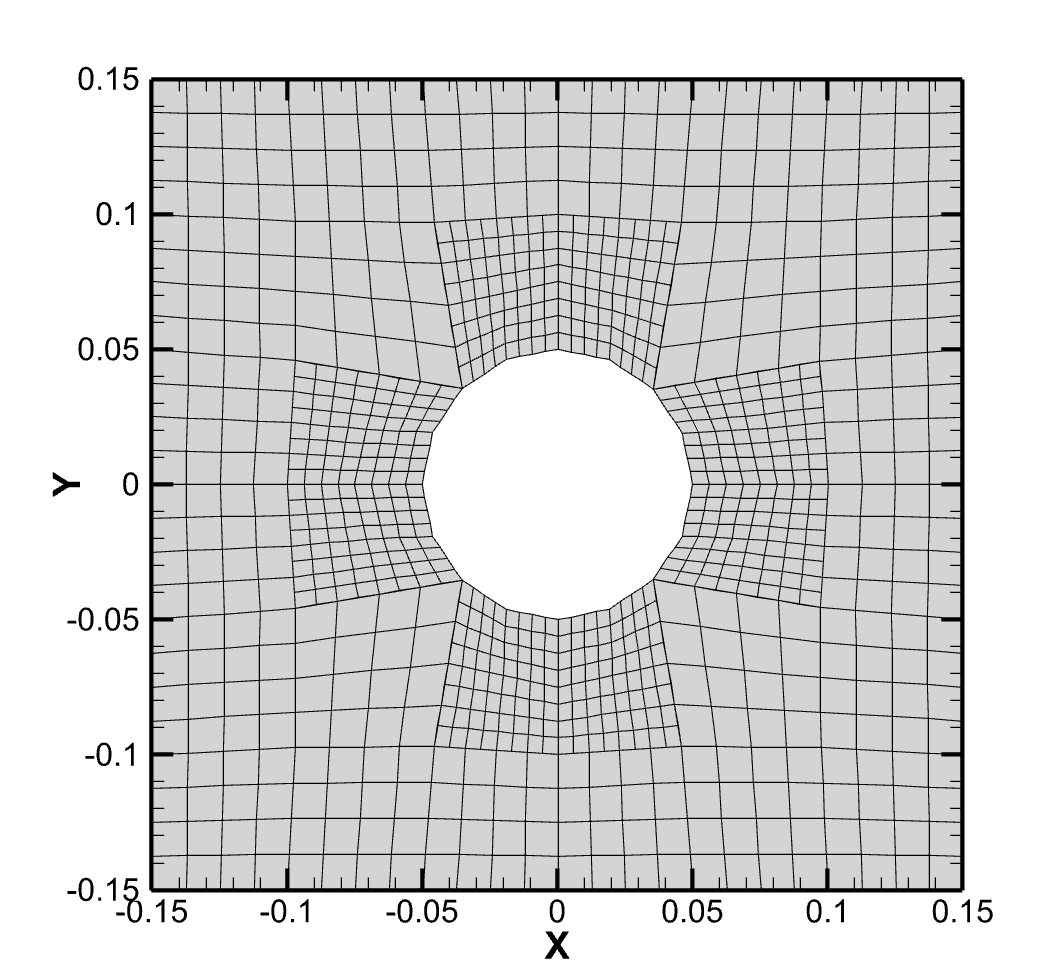} 
      \caption*{(b) with block-based correction}
    \end{minipage}
    \hspace{0.05\textwidth}
    \begin{minipage}[b]{0.45\textwidth}
      \centering
      \includegraphics[height=6cm]{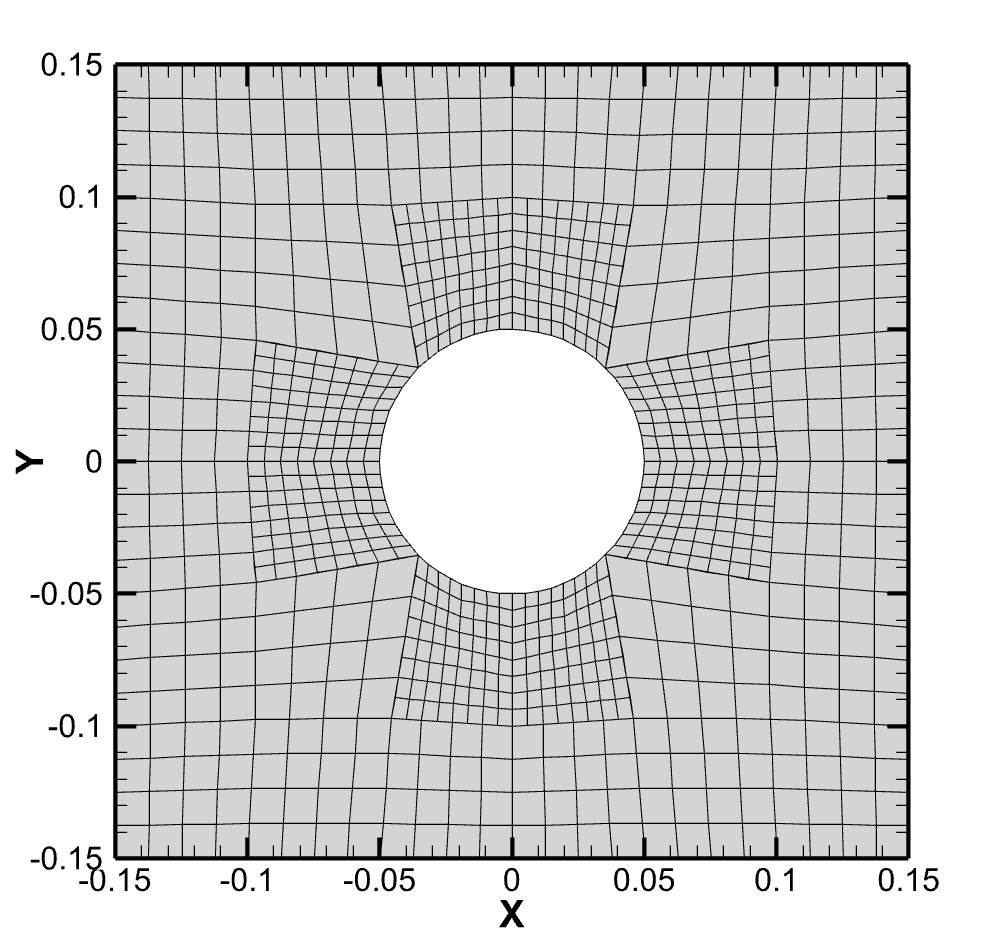} 
      \caption*{(c) with cell-based correction}
    \end{minipage}
  \caption{\label{fig3} Comparison of block-based and cell-based nodal correction methods for curved boundaries. } 
\end{figure}
In the numerical examples presented in this study, 
a key objective is to establish an appropriate adaptive criterion for AMR in regions 
exhibiting vortex structures. 
To achieve this, we define a dimensionless vorticity parameter in the z-direction:$\omega^*_z=\frac
{\omega_zd}{2|\mathbf{V}_{\infty}|}$, where $\omega_z$ represents the physical vorticity component in the z-direction, 
$d$ denotes the cylinder diameter, and $\mathbf{V}_{\infty}$ corresponds to the free-stream velocity. 
The adaptive strategy is implemented through two threshold values:
$\omega_\mathrm{min}$ and $\omega_\mathrm{max}$. Local mesh refinement is triggered when $\omega^*_z>\omega_\mathrm{max}$, 
and mesh coarsening is applied when $\omega^*_z<\omega_\mathrm{min}$. 
This dual-threshold approach ensures concentrated computational resolution 
in high-vorticity regions while maintaining efficiency through coarsening 
in low-vorticity regions. 

The high-precision nature of DG methods presents a challenge 
in geometrically complex regions: excessively dense grids incur prohibitive computational costs, 
while overly coarse grids compromise boundary representation fidelity. 
To resolve this problem, Kong et al. \cite{kong2023numerical} (2023) proposed an AMR- driven 
block node correction method for curved boundaries. 
This approach iteratively adjusts nodes of AMR blocks intersecting curved surfaces 
to progressively converge toward theoretical boundary profiles. 
Building upon this foundation, this study propose an enhanced cell node correction method 
that significantly improves the geometric approximation capability of quadrilateral 
elements for curved boundaries through AMR-driven refinement. 

The block node correction method repositions nodes of all refined blocks 
adjacent to curved boundaries to their theoretical positions. 
Our cell node correction method extends this framework by further 
optimizing nodes of individual cells along block edges. 
As an illustrative example using a circular geometry with radius $R=0. 05$, 
Fig.~\ref{fig3} demonstrates the comparative corrective performance between the 
block-based nodal correction method proposed by Kong et al. 
and the cell-based method developed in this work. 

\subsection{\label{sec2.5}Ghost-node boundary condition treatment for moving mesh}
In CFD, appropriate boundary condition implementation constitutes a critical methodological
 consideration. Improper boundary treatments may induce significant computational errors 
 or even divergence. 
 To address this challenge, this study develops a moving boundary condition strategy 
 through a two-tiered approach: 
 1) Domain Coupling: Establishes correspondence between external
  ghost points $(\cdot)_R$ and internal physical Gauss points $(\cdot)_L$. 
  2) ghost Gauss Point Extension: Prescribes values to an extended layer of ghost Gauss points 
  beyond the computational domain boundary. 
  All numerical integrations employ Gauss-Lobatto quadrature points (Fig.~\ref{fig4}), 
  selected for their enhanced numerical stability and spectral accuracy characteristics 
  at domain boundaries. This section introduces three types of moving-wall boundary 
  conditions implemented through the ghost node method:
  \begin{figure}[h]
    \includegraphics[width=0.45\textwidth]{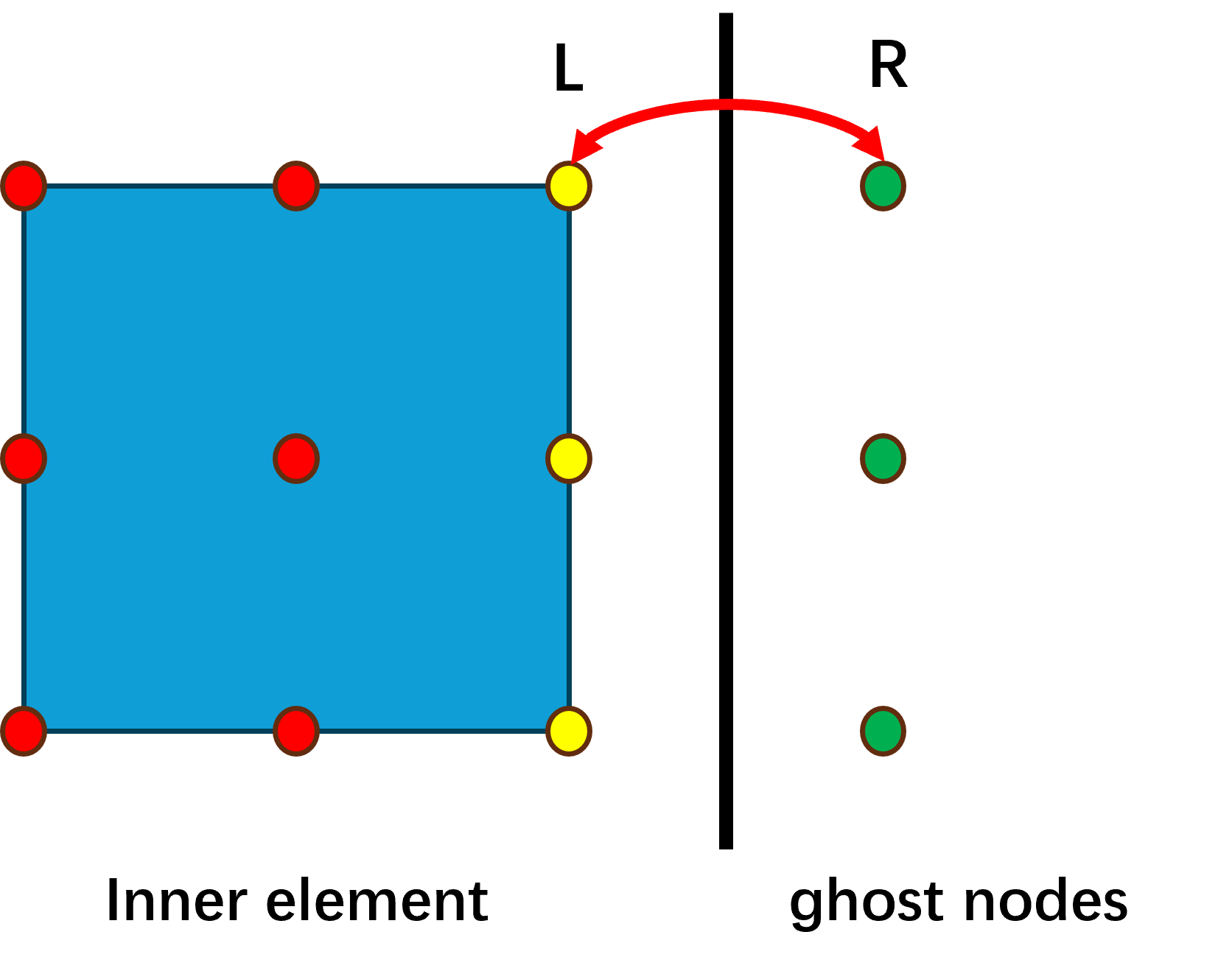}
    \caption{\label{fig4} Schematic of ghost-node boundary condition treatment. }
  \end{figure}

  (1) Adiabatic no-slip boundary condition for moving surfaces. 

  The fundamental mathematical formulation of the adiabatic no-slip boundary condition is expressed as:
  \begin{equation}
  \begin{aligned}
   \mathbf{V}=\mathbf{V}_{wall}&=\mathbf{V}_{mesh}, \\
    \frac{\partial T}{\partial \mathbf{n}}&=0. 
  \end{aligned}
  \label{eq:18}
  \end{equation}

  Following the principles of boundary layer theory, the following relationships can be derived:
  \begin{equation}
  \begin{aligned}
    p_R &= p_L , \\
    T_R &= T_L , \\
    e_R &= e_L , \\
    \rho_R &= \rho_L , \\
    \mathbf{V}_R &= -\mathbf{V}_L+2\mathbf{V}_{mesh}. 
   \end{aligned}  
   \label{eq:19}
   \end{equation}

  (2) Isothermal no-slip boundary condition for moving surfaces. 

  The isothermal no-slip boundary condition differs from the adiabatic no-slip 
  condition in that the boundary temperature is explicitly prescribed as follows:
  \begin{equation}
  \begin{aligned}
    \mathbf{V}=\mathbf{V}_{wall}&=\mathbf{V}_{mesh}, \\
     T &= T_{wall}. 
   \end{aligned}
   \label{eq:20}
    \end{equation}

   Following the principles of boundary layer theory and Tammann eos, 
   the following relationships can be derived:
   \begin{equation}
   \begin{aligned}
    T_R &= T_{wall} , \\
    e_R &= C_vT_R, \\
    p_R &= p_L, \\
    \rho_R &= \frac{p_R+\gamma P_w}{e_R(\gamma -1)}, \\
    \rho _L\mathbf{V}_L +\rho _R\mathbf{V}_R &= (\rho _L+\rho _R)\mathbf{V}_{mesh}. 
   \end{aligned}  
   \label{eq:21}
  \end{equation}

  (3) Dynamic jet boundary condition. 

  When simulating jet flow for vortex-induced vibration suppression 
  (see Section~\ref{SJs-VIV} for details), 
  dynamic jet boundary conditions must be imposed on the jet orifices:
  \begin{equation}
    \mathbf{V}_R=\mathbf{V}_{wall}+\mathbf{V}_{sj}. 
    \label{eq:22}
  \end{equation}
  \subsection{\label{sec2.6}Governing equation for the oscillating cylinder and computational methods for FSI}
  Using the transverse VIV of a circular cylinder as a benchmark case, 
  we present the structural governing equations, moving mesh FSI computational framework, 
  and FSI time-stepping scheme for FSI analysis. 
  The structural governing equations for transverse VIV of a circular cylinder \cite{schulz1998unsteady} 
  can be formulated as:
  \begin{equation}
    m\ddot{y}+c\dot{y}+ky={F}_{l}(t), 
    \label{eq:23}
  \end{equation}  
  in which ${F}_{l}(t)$ denotes the time-varying lift acting on the cylinder, 
  $y$ represents the instantaneous position of the rigid cylinder, $\dot{y}$ and $\ddot{y}$
  denote its velocity and acceleration, respectively. The parameters $m$ , $c$ and $k$ correspond to the cylinder mass, 
  structural damping coefficient, and stiffness coefficient. To nondimensionalize the structural governing Eq.~(\ref{eq:23}), 
  we introduce the following dimensionless variables and parameters:
  \begin{equation}
  \begin{aligned}
    & Y=\frac{y}{d}, T=\frac{t}{d/{{u}_{\infty }}}, \\ 
   & {{C}_{l}}=\frac{2{{F}_{l}}}{{{\rho }_{\infty }}u_{\infty }^{2}d}, {{C}_{d}}=\frac{2{{F}_{d}}}{{{\rho }_{\infty }}u_{\infty }^{2}d}, St=\frac{{{f}_{v}}d}{{{u}_{\infty }}} , \\
   & {{m}^{*}}=\frac{m}{{{\rho }_{\infty }}{{d}^{2}}}, {{c}^{*}}=\frac{c}{2\sqrt{km}} , \\ 
   & {{U}_{R}}=\frac{{{u}_{\infty }}}{{{f}_{n}}d}, {{f}_{n}}=\frac{1}{2\pi }\sqrt{\frac{k}{m}} , \\
   & \mathrm{Re}=\frac{{{\rho }_{\infty }}{{u}_{\infty }}d}{\mu }, \mathrm{Ma}=\frac{{{u}_{\infty }}}{{{c}_{\infty }}} , 
  \end{aligned}
  \label{eq:24}
  \end{equation}
  where \( Y \) is the dimensionless displacement, and \( T \) is the dimensionless time. 
\( \rho_{\infty} \) denotes the free-stream fluid density, 
\( C_l \) and \( C_d \) represent the lift and drag coefficients, respectively, 
\( f_v \) and \( St \) are the vortex shedding frequency and Strouhal number. 
The mass ratio \( m^* \) characterizes the hydrodynamic inertia of the cylinder, 
and \( c^* \) is the dimensionless damping coefficient. 
The reduced velocity \( U_R \) quantifies the stiffness of the structural system:  
 larger \( U_R \) corresponds to smaller natural frequency \( f_n \), 
indicating reduced structural stiffness. 
Here, \( f_n \) denotes the natural frequency of the structural system. 
Additionally, \( \mathrm{Re} \) (Reynolds number) and \( \mathrm{Ma} \) (Mach number) characterize the flow regime, 
and \( c_{\infty} \) is the local speed of sound in the free-stream fluid. 
The formulations for lift and drag calculations are given in Eq.~(\ref{eq:29}):
\begin{equation}
  \begin{aligned}
    & {{F}_{d}}={{F}_{dp}}+{{F}_{df}}=\oint{-p{{n}_{x}}}dl+\oint{{{\tau }_{xx}}{{n}_{x}}+{{\tau }_{yx}}{{n}_{y}}}dl , \\ 
    & {{F}_{l}}={{F}_{lp}}+{{F}_{lf}}=\oint{-p{{n}_{y}}}dl+\oint{{{\tau }_{xy}}{{n}_{x}}+{{\tau }_{yy}}{{n}_{y}}}dl , 
  \end{aligned}
  \label{eq:29} 
\end{equation}
here, \( \mathbf{n} = \left( n_x, n_y \right) \) denotes the unit outward 
normal vector of the cylinder. 

The structural governing Eq.~(\ref{eq:23}) can be nondimensionalized 
to yield the following dimensionless form:
\begin{equation}
  \frac{{{d}^{2}}Y}{d{{T}^{2}}}+(\frac{4\pi {{c}^{*}}}{{{U}_{R}}})\frac{dY}{dT}+
  \left( \frac{4{{\pi }^{2}}}{U_{R}^{2}} \right)Y=\frac{{{C}_{l}}}{2{{m}^{*}}}
  \label{eq:25}   
\end{equation}

\begin{figure}
        \centering
        \includegraphics[width=0.45\textwidth]{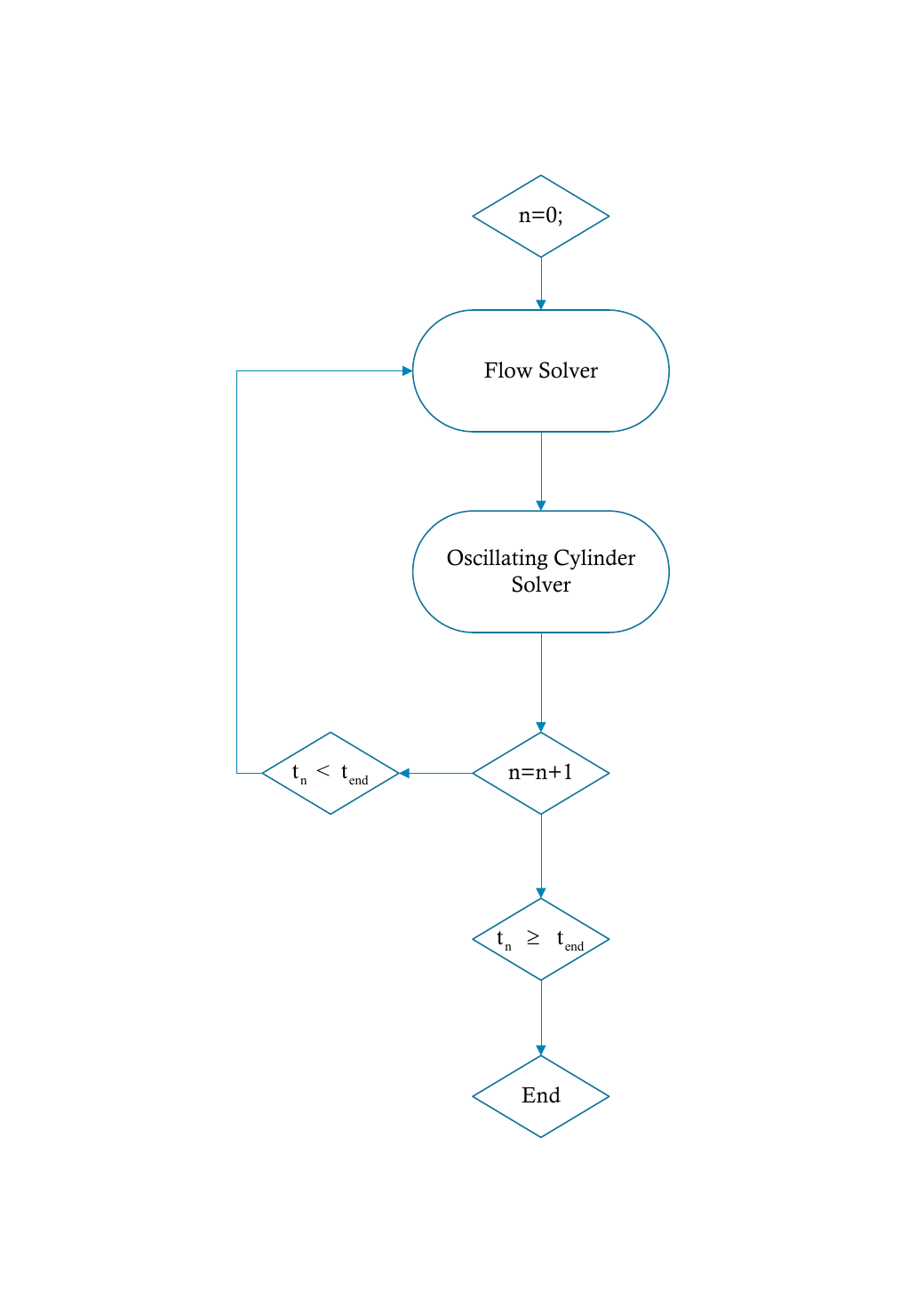}
        \caption{Schematic of computational procedure for FSI. }
        \label{fig5}
    \end{figure}

The computational framework of the moving mesh FSI method for transverse VIV of a circular cylinder
 is illustrated in Fig.~\ref{fig5}. The specific computational procedure is as follows:
 \begin{enumerate}
  \item \textbf{Initialization:}  
  Assume the acceleration, velocity, and displacement of the cylinder at time \( t_n \) are known:  
  $\ddot{y}^{t_n}, \quad \dot{y}^{t_n}, \quad y^{t_n}$. 
  
  \item \textbf{Time-Step Update:}  
  Update the velocity and displacement at \( t_{n+1} \) using the values at \( t_n \):  
  $\dot{y}^{t_{n+1}}, \quad y^{t_{n+1}}$. 
  
  \item \textbf{Fluid-Solid Coupling:}  
  \begin{itemize}
      \item Impose the mesh velocity \( \mathbf{V}_{mesh} = (0, \dot{y}^{t_n}) \) into Eq.~(\ref{eq:8})
      to update the governing equations and cylinder boundary conditions. 
      \item Solve the flow field using the IPDG method to obtain 
      the updated flow variables at \( t_{n+1} \):  
      $\mathrm{\mathbf{U}}_{t_{n+1}}$. 
      \item Compute the lift at \( t_{n+1} \):  
      $F_l\left(\mathrm{\mathbf{U}}_{t_{n+1}}\right)$. 
  \end{itemize}
  
  \item \textbf{Oscillating cylinder Dynamics:}  
  Calculate the updated acceleration at \( t_{n+1} \) using Eq.~(\ref{eq:23}):  
  $\ddot{y}^{t_{n+1}}$. 
\end{enumerate}
Through this iterative process, the acceleration, 
velocity, and displacement at \( t_n \)—denoted as \( \ddot{y}^{t_n}, 
\dot{y}^{t_n}, y^{t_n} \)—are advanced to \( t_{n+1} \), yielding \( \ddot{y}^{t_{n+1}}, 
\dot{y}^{t_{n+1}}, y^{t_{n+1}} \). 

In the temporal advancement of the moving mesh FSI framework, 
this work employs a third-order TVD-Runge-Kutta method to concurrently 
advance both the fluid and structural subsystems. Let $\mathbf{Z}$ 
denote the coupled fluid-structure state vector undergoing time integration, defined as:
\begin{equation}
  \mathbf{Z} = \left[
    \begin{array}{c}
      \mathrm{\mathbf{U}}\\
      y\\
      \dot{y}
    \end{array}
    \right]
    \label{eq:26}
\end{equation}

The time advancement of the coupled fluid-structure state vector $\mathbf{Z}$ is shown:
\begin{equation}
\begin{aligned}
  & \mathbf{Z}^{(1)} = \mathbf{Z}^{t_n} + \Delta t \dot{\mathbf{Z}}^{t_n} \\
  & \ddot{y}^{(1)} = \left( F_l^{(1)}\left( \tilde{\mathbf{U}}^{(1)} \right) - c \dot{y}^{(1)} - k y^{(1)} \right) \big/ m \\
  & \mathbf{Z}^{(2)} = \frac{3}{4} \mathbf{Z}^{t_n} + \frac{1}{4} \mathbf{Z}^{(1)} + \frac{1}{4} \Delta t \dot{\mathbf{Z}}^{(1)} \\
  & \ddot{y}^{(2)} = \left( F_l^{(2)}\left( \tilde{\mathbf{U}}^{(2)} \right) - c \dot{y}^{(2)} - k y^{(2)} \right) \big/ m \\ 
  & \mathbf{Z}^{t_{n+1}} = \frac{1}{3} \mathbf{Z}^{t_n} + \frac{2}{3} \mathbf{Z}^{(2)} + \frac{2}{3} \Delta t \dot{\mathbf{Z}}^{(2)} \\ 
  & \ddot{y}^{t_{n+1}} = \left( F_l^{(3)}\left( \tilde{\mathbf{U}}^{(3)} \right) - c \dot{y}^{t_{n+1}} - k y^{t_{n+1}} \right) \big/ m 
\end{aligned}
\label{eq:27}
\end{equation}

\section{NUMERICAL VERIFICATION}
\subsection{\label{coue}Couette flow}
When the mesh remains stationary ($\mathbf{V}_{mesh}\equiv \mathbf{0}$), 
Eq.~(\ref{eq:8}) reduces to Eq.~(\ref{eq:1}). 
The Couette flow, serving as an analytical solution to the compressible NS equations, 
is frequently employed to validate the order-of-accuracy of NS solvers. 
The physical configuration comprises viscous fluid motion between two parallel plates with 
the following boundary specifications:the lower plate is fixed with constant temperature $T_0$, 
the upper plate has a velocity $V_1$ with constant temperature $T_1$, 
the gap height between the plates is $H$. 
Under constant dynamic viscosity $\mu$, the analytical solution \cite{sun2007high} for this problem is given by:
\begin{equation}
  \begin{aligned}
  &u = \frac{V_1}{H}y, v=0, \\
  &T = T_0 + \frac{T_1-T_0}{H}y+\frac{\mu V_1^2}{2k}\cdot \frac{y}{H}(1-\frac{y}{H}), \\
  &p = constant, \rho = \frac{p}{RT}. 
  \end{aligned}
  \label{eq:28}
\end{equation}

In this case, $V_1 = 1. 0, H =1. 0, T_0=0. 8, T_1=0. 85, \mu = 0. 01$. 
The L2-norm error convergence and accuracy of density computed by solving the NS equations 
using IPDG methods of different orders are summarized in Table~\ref{tab:1}. 
Here, IP(2), IP(3), and IP(4) denote the Interior Penalty Discontinuous Galerkin 
schemes with second-order, third-order, and fourth-order accuracy, respectively. 
\begin{table*}
  \centering
  \caption{Convergence and accuracy of density computed by solving the NS equations using IPDG methods of different orders. }
  \begin{ruledtabular}
  \begin{tabular}{ccccccc}
    grid &\multicolumn{2}{c}{IP(2)}&\multicolumn{2}{c}{IP(3)}&\multicolumn{2}{c}{IP(4)}\\ \hline
    & L2 error&Order&L2 error&Order&L2 error&Order\\
    2$\times$2&5. 18e-3& &6. 8e-5& &1. 64e-6&  \\ 
    4$\times$4&1. 28e-3&2. 0168&8. 7e-6&2. 9664&8. 52e-8&4. 2667\\
    8$\times$8&3. 2e-4&2. 00&1. 02e-6&3. 0924&4. 35e-9&4. 2918\\
  \end{tabular}
  \end{ruledtabular}
  \label{tab:1}
\end{table*}

As can be seen from Table~\ref{tab:1}, 
even on the same 2$\times$2 grid, 
as the IPDG order increases from 2 to 4, 
the error magnitude decreases from 1e-3 to 1e-5 and further to 1e-6. 
This demonstrates that elevating the IPDG order can significantly improve the 
accuracy of flow field simulations, 
thereby validating the high-order accuracy characteristics of the solver 
developed in this study. 
\subsection{\label{cy flow}Flow over a cylinder}
\begin{figure}
  \includegraphics[width=0.45\textwidth]{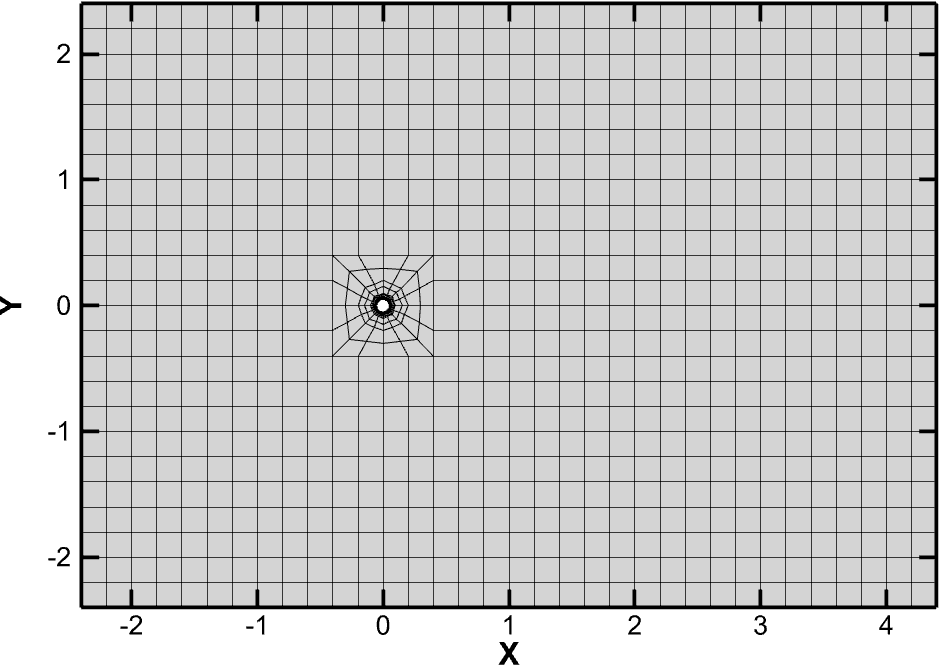}
  \caption{\label{fig6} The initial block partitioning of the computational domain. }
\end{figure}

\begin{figure}
  \includegraphics[width=0.45\textwidth]{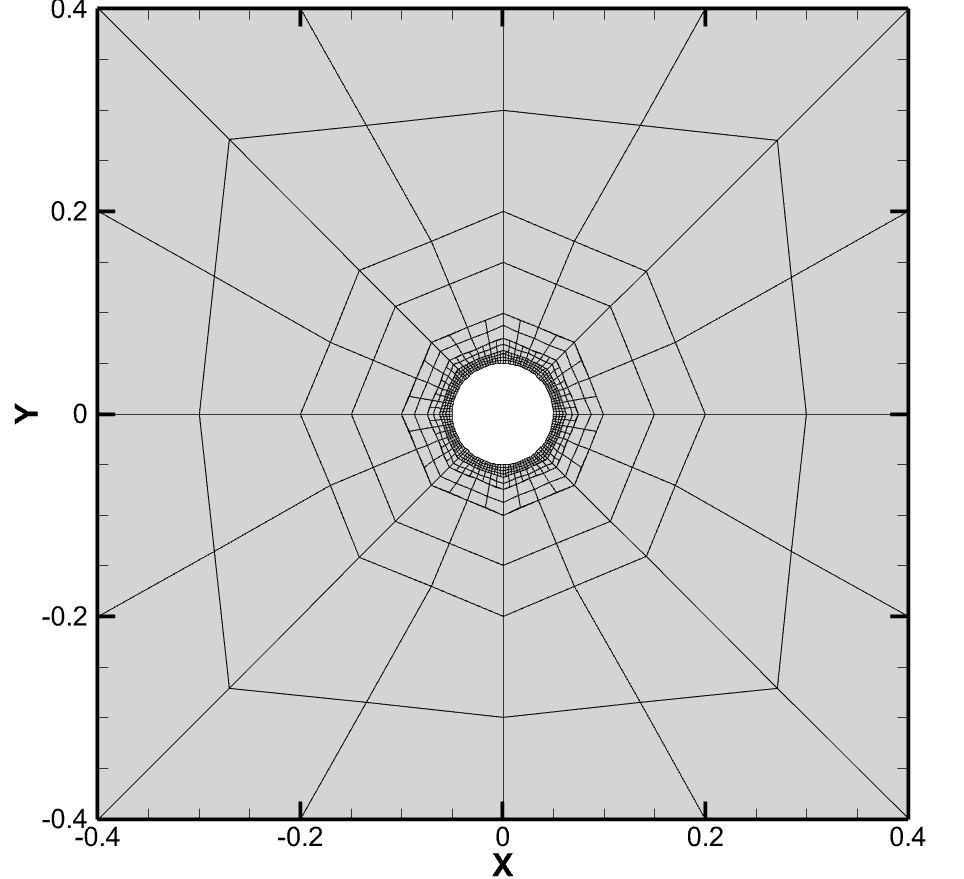}
  \caption{\label{fig7} The local initial mesh near the cylinder. }
\end{figure}

This section presents a two-dimensional unsteady simulation 
of cylinder flow. The cylinder diameter is set as $d=0. 1$. 
The freestream conditions are Mach number $\mathrm{Ma}=0. 2$ and 
Reynolds number $\mathrm{Re}=150$. The computational domain spans [-2. 4, 4. 4]$\times$[-2. 4, 2. 4] 
with the cylinder center located at (0, 0). 
An adiabatic no-slip boundary condition is applied to the cylinder wall, 
where the cylindrical surface is constructed using the cell-based node correction method 
proposed in this study. Non-reflective boundary conditions \cite{engquist1977absorbing} 
are implemented at all far-field boundaries. 
To demonstrate the superiority of the RKDG AMR framework, 
dynamic AMR technology is employed to capture vortex structures. 
The initial block partitioning of the computational domain and the local initial mesh 
near the cylinder are shown in Fig.~\ref{fig6} and Fig.~\ref{fig7}, respectively. 

\begin{figure}
  \includegraphics[width=0.45\textwidth]{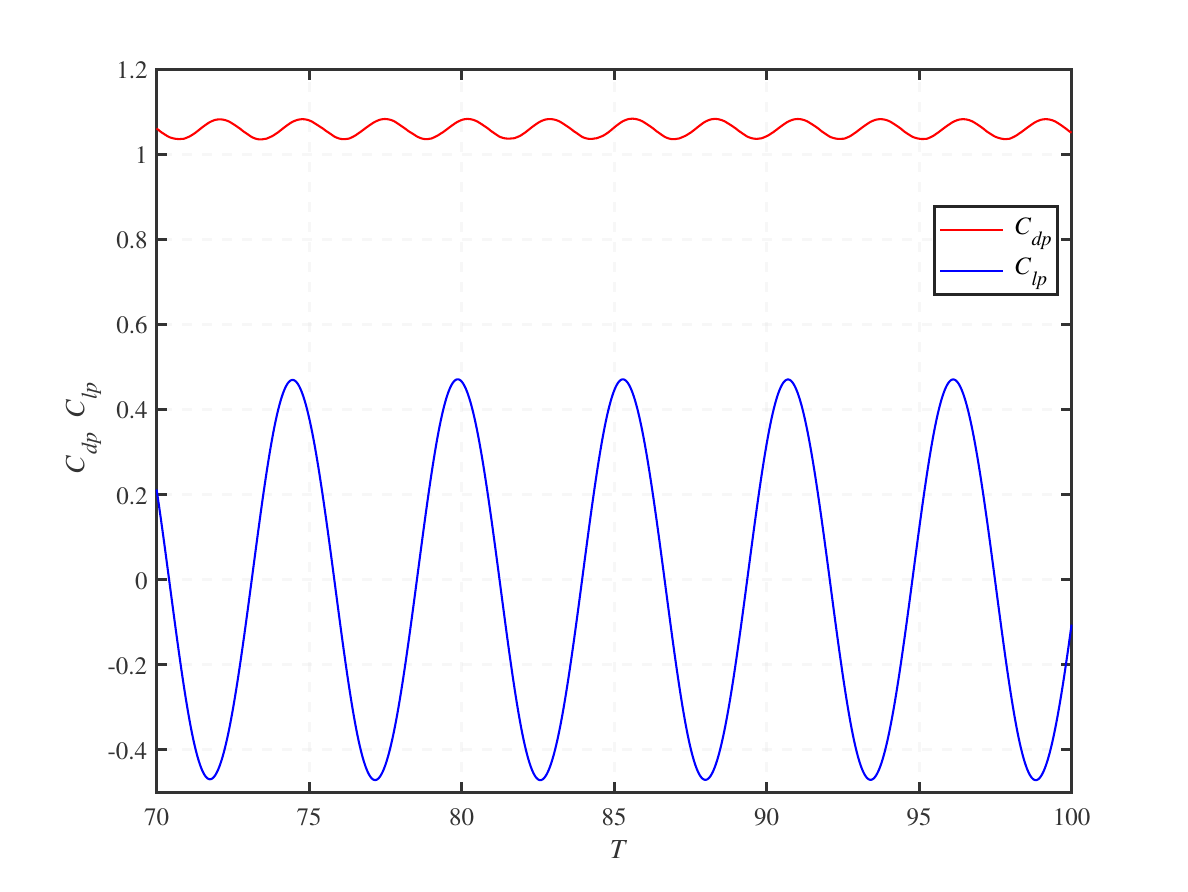}
  \caption{\label{fig8} The time evolution of the pressure lift and pressure drag coefficients computed by IP(4). }
\end{figure}

\begin{figure}
  \includegraphics[width=0.45\textwidth]{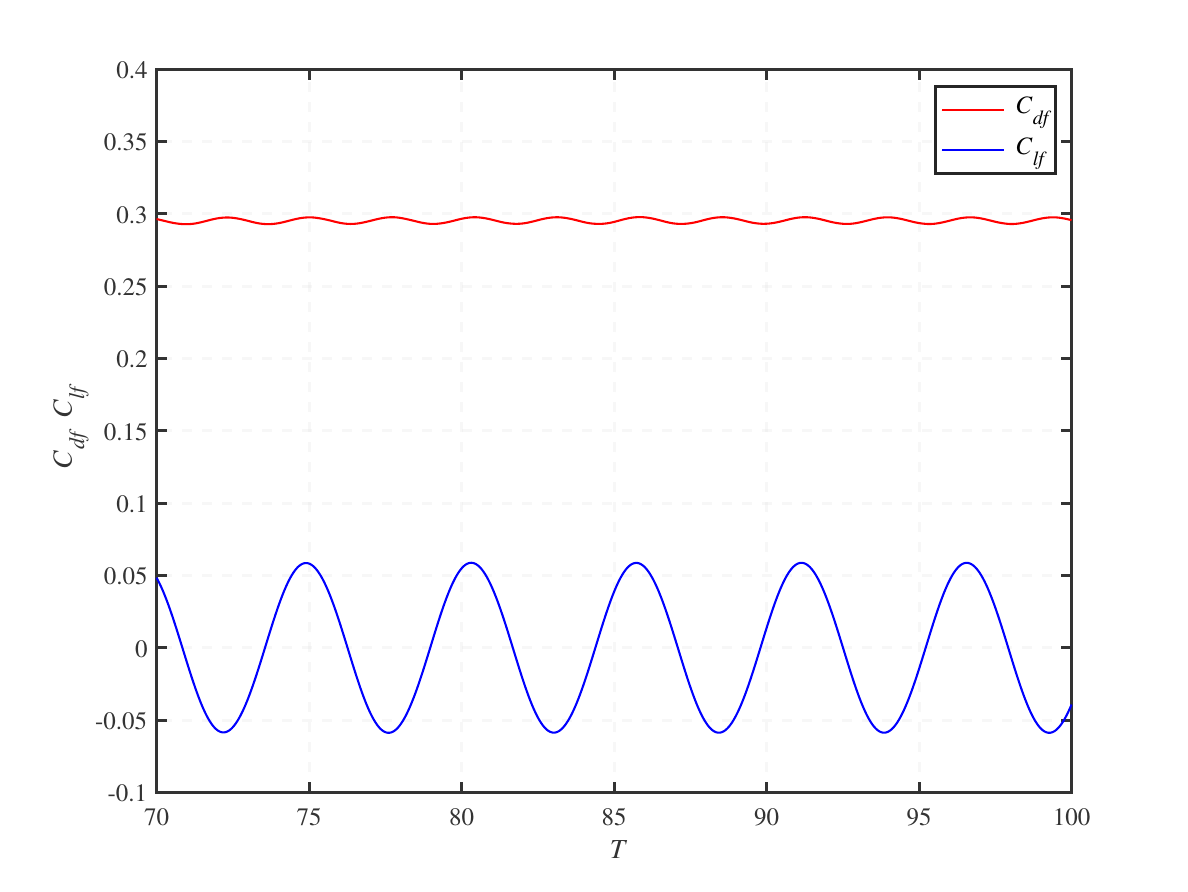}
  \caption{\label{fig9} The time evolution of the viscous lift and viscous drag coefficients computed by IP(4). }
\end{figure}

\begin{figure}  
  \includegraphics[width=0.45\textwidth]{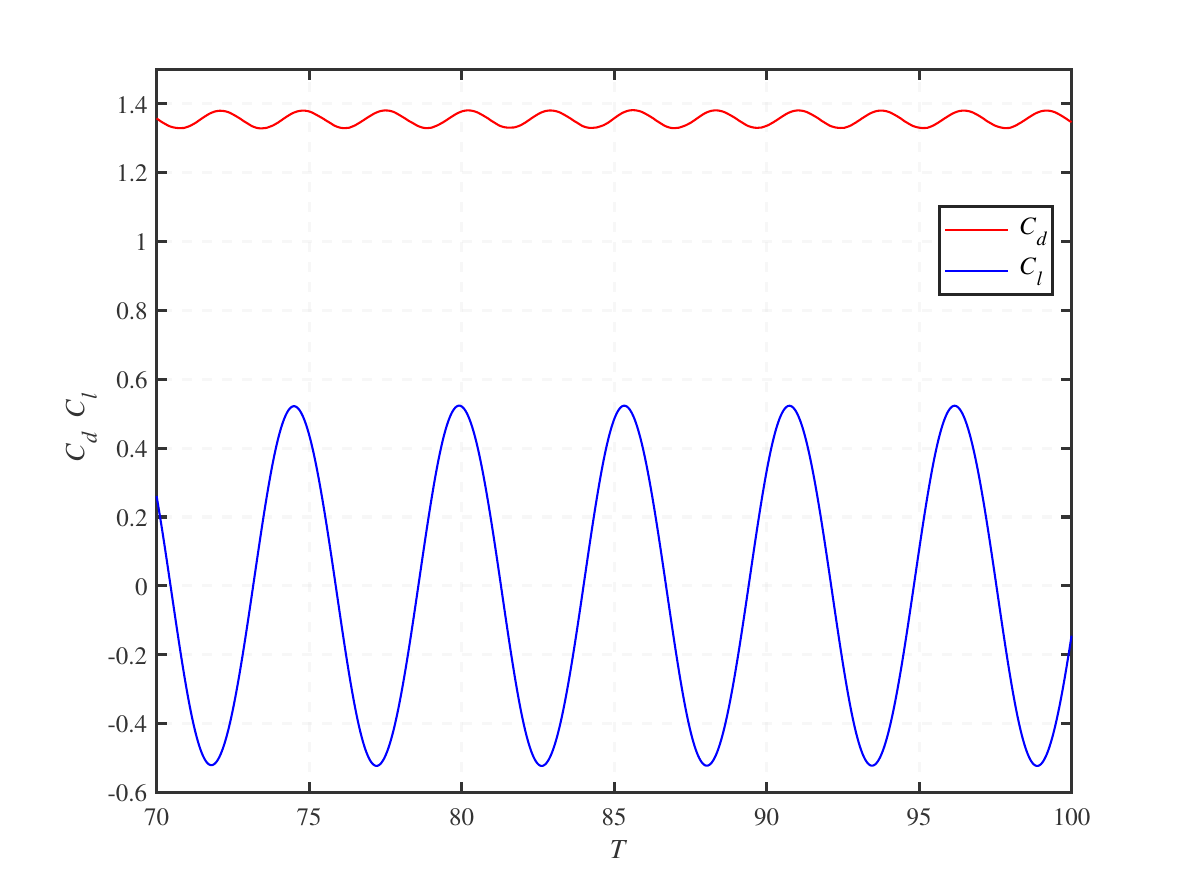}
  \caption{\label{fig10} The time evolution of the lift and drag coefficients computed by IP(4). }
\end{figure}

In this section, 
numerical solutions of the compressible NS equations
(where Eq.~(\ref{eq:8}) reduces to the NS equations under stationary grid conditions) 
are obtained using the IP(2), IP(3), and IP(4) schemes, respectively. 
The integration domains for the pressure lift \( F_{lp} \), 
viscous lift \( F_{lf} \), pressure drag \( F_{dp} \), 
and viscous drag \( F_{df} \) are all defined over the cylinder boundary. 
The computed lift coefficient, drag coefficient, and Strouhal number 
are compared against those reported by Zhang \cite{zhang2014class} and 
Müller et al. \cite{muller2008high}, 
as summarized in Table~\ref{tab:2}. 
\begin{table*}
  \centering
  \caption{Comparison of lift and drag coefficients and Strouhal numbers. }
  \begin{ruledtabular}
  \begin{tabular}{ccccc}
    &Time-averaged $C_d$&$\triangle C_d$ peak to peak&$\triangle C_l$ peak to peak&Strouhal number\\\hline
    IP(2)	&1. 33	&0. 0497	&1. 016	&0. 187\\
    IP(3)	&1. 3503&	0. 0518	&1. 04&	0. 1845\\
    IP(4)	&1. 3547&	0. 0512	&1. 0456&	0. 1847\\
    Zhang&1. 348&	0. 0519&	1. 048	&0. 184\\
    Müller&1. 34&	0. 05228&	1. 0406&	0. 183\\
  \end{tabular}
  \end{ruledtabular}
  \label{tab:2}
\end{table*}
The time evolution of the pressure lift and pressure drag coefficients computed by IP(4) are shown in Fig.~\ref{fig8}. 
The time evolution of the viscous lift and viscous drag coefficients computed by IP(4) are shown in Fig.~\ref{fig9}. 
The time evolution of the lift and drag coefficients computed by IP(4) are shown in Fig.~\ref{fig10}. 

\begin{figure}
  \begin{minipage}[b]{0.45\textwidth}
    \centering
    \includegraphics[height=4.5cm]{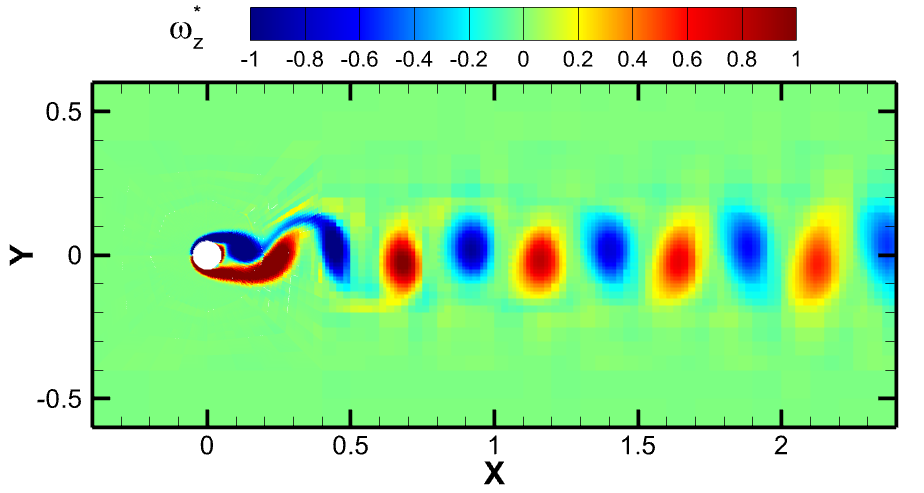} 
    \caption*{(a) IP(2)}
  \end{minipage}
  \begin{minipage}[b]{0.45\textwidth}
    \centering
    \includegraphics[height=4.5cm]{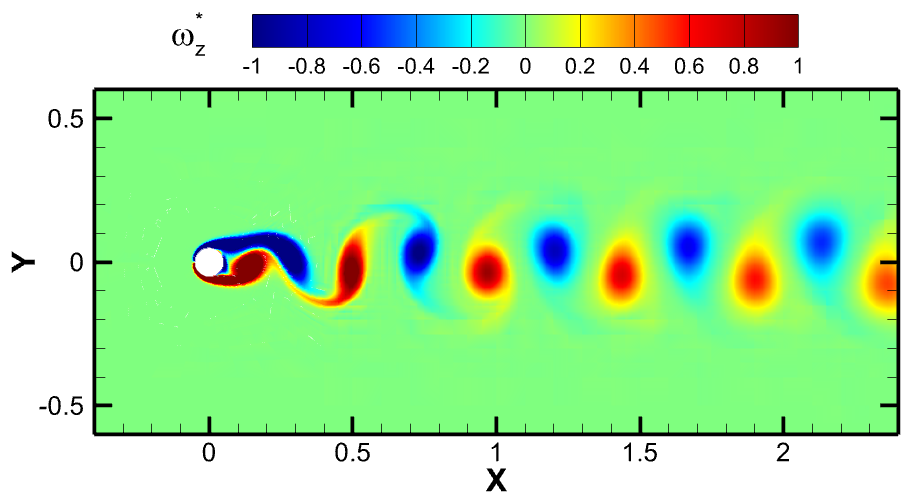} 
    \caption*{(b) IP(3)}
  \end{minipage}
  \begin{minipage}[b]{0.45\textwidth}
    \centering
    \includegraphics[height=4.5cm]{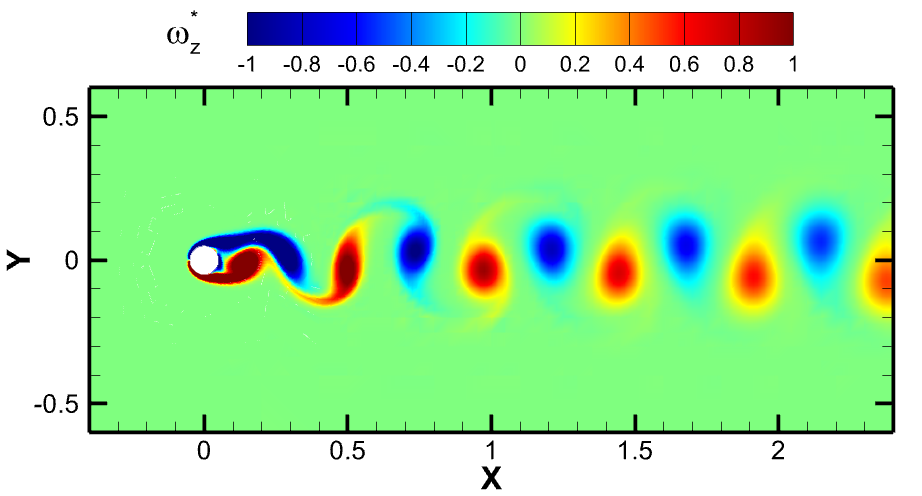} 
    \caption*{(c) IP(4)}
  \end{minipage}  
  \caption{\label{fig11} The dimensionless vorticity $\omega_z^*$ contour plots computed using IP(2), IP(3), and IP(4) at $T$=100. }
\end{figure}

Figure.~\ref{fig11} compares the dimensionless vorticity $\omega_z^*$ contours simulated 
using IP(2), IP(3), and IP(4) methods at an identical time instance $T$=100. 
The lower-order IP(2) scheme exhibits significant distortion in flow field 
evolution due to its reduced accuracy and amplified numerical dissipation. 
In contrast, the higher-order IP(3) and IP(4) formulations demonstrate 
remarkable convergence, with their vorticity patterns showing negligible differences. 

As clearly shown in Fig.~\ref{fig11}, 
the $\omega_z^*$ contour plots obtained with the IP(2) 
scheme exhibit relatively low resolution. 
However, when the solution accuracy is elevated to IP(3) and IP(4) 
levels, a significant enhancement in the resolution of $\omega_z^*$ 
contours is achieved, yielding high-fidelity flow field visualizations. 
This demonstrates that for the same coarse grid configuration, 
increasing the IPDG order can effectively 
improve solution precision, thereby fully validating 
the superior capability of high-order IPDG methods 
in resolving complex fluid dynamic phenomena. 

\begin{figure}
  \includegraphics[height=4.5cm]{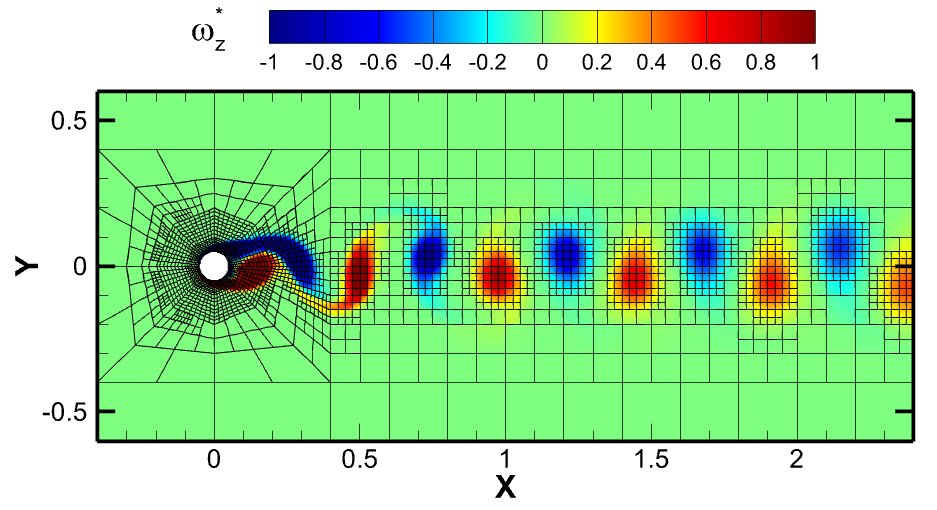}
  \caption{\label{fig:12} The dimensionless vorticity $\omega_z^*$ contour plots 
  and AMR block partitioning computed using IP(4). }
\end{figure}

The dimensionless vorticity $\omega_z^*$ contour plots 
and AMR block partitioning computed using the IP(4) 
method are shown in Figure~\ref{fig:12}. 
Figure~\ref{fig:12} demonstrates that the blocks are dynamically 
refined to the maximum adaptive level $A_{max}$ in 
regions with sufficiently large $\omega_z^*$ values, 
whereas grid is maintained in areas with diminished $\omega_z^*$ magnitudes. 
This adaptive refinement strategy achieves substantial computational efficiency gains 
while preserving solution accuracy, 
thereby validating the superior performance of the AMR technology used in this study. 
\subsection{\label{SJs-VIV}Synthetic jets flow control for VIV}
To balance computational accuracy and efficiency, 
the IP(3) scheme is employed to solve Eq.~(\ref{eq:8}). 
Subsequently, six direct numerical simulations (DNS) 
of transverse VIV benchmark cases are simulated using
IP(3) within the RKDG AMR framework, establishing the proposed FSI approach's accuracy. 
Furthermore, the FSI approach is employed to 
investigate active flow control SJs for VIV suppression \cite{wang2024effect, wang2024numerical, wang2016control}, 
with particular emphasis on the correlation between 
SJs actuation frequency and suppression performance. 

\begin{figure}
  \includegraphics[width=0.45\textwidth]{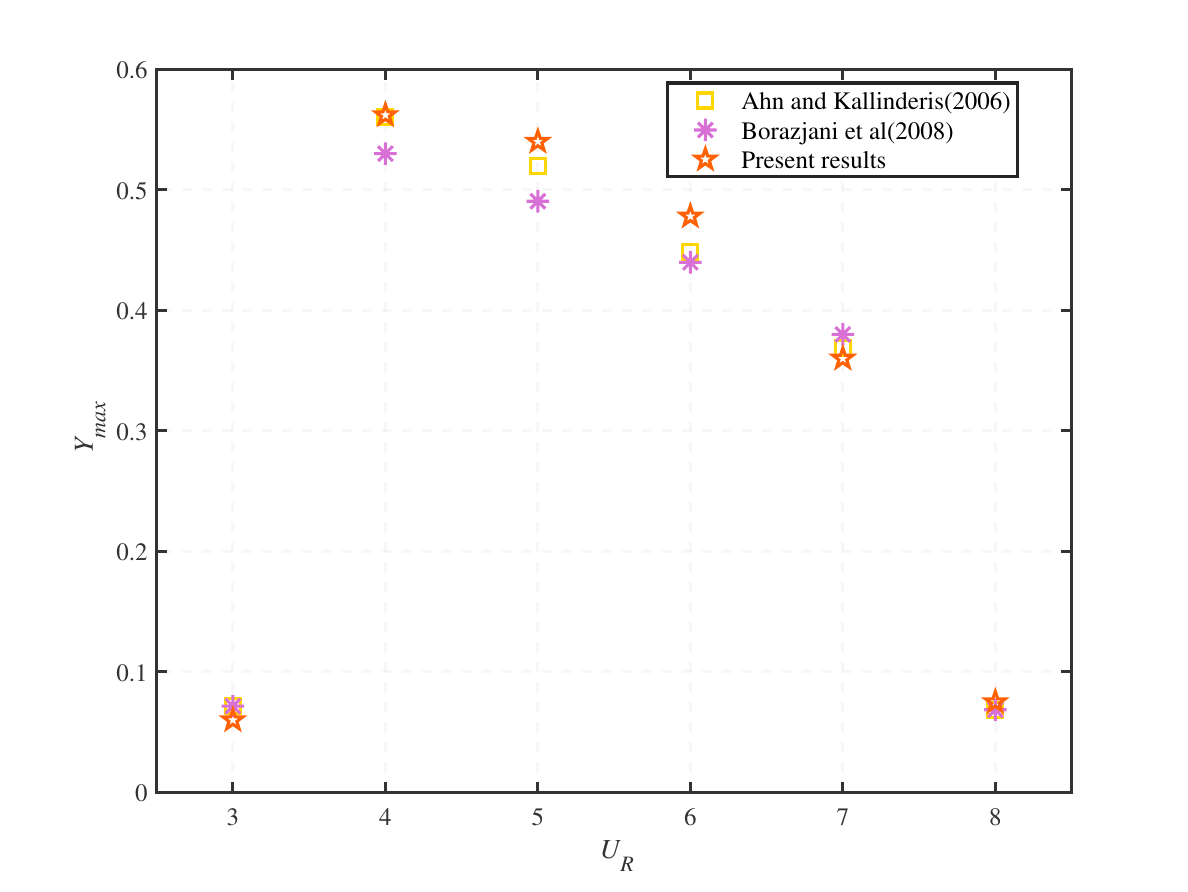}
  \caption{\label{fig:13} The maximum dimensionless displacement \( Y_{\max} \) computed by IP(3). }
\end{figure}

All numerical cases in this section 
employ the identical computational domain 
as specified in Section~\ref{cy flow}, 
with the cylindrical boundary prescribed as a 
transient-moving adiabatic no-slip boundary condition. 
The parameters for the transverse VIV benchmark cases are set as:
\[
\mathrm{Re} = 150, \quad m^* = 2, \quad c^* = 0. 0, \quad U_R = 3. 0 \text{--} 8. 0
\]
with a Mach number \( \mathrm{Ma} = 0. 085 < 0. 2 \), 
satisfying the incompressibility condition. 
The computed results of the maximum dimensionless displacement 
\( Y_{\max} \) are shown in Fig.~\ref{fig:13}. 

Fig.~\ref{fig:13} presents the scatter plot of vibration amplitude versus 
reduced velocity $U_R$ for transverse VIV of a single cylinder. 
To validate the accuracy of the proposed FSI approach, 
comparative data from Ahn \cite{AHN2006671} and Borazjani et al. \cite{borazjani2008curvilinear} 
are included. As depicted in the figure, 
our results exhibit excellent agreement with these benchmark studies. 
Key observations reveal distinct regime transitions: 
1) $U_R$ = 3: The cylinder undergoes low-amplitude periodic oscillations about its axis. 
2) $U_R$ transition 3 to 4: A dramatic amplitude escalation occurs as the vortex 
shedding frequency approaches the structural natural frequency, 
triggering fluidelastic resonance. 
3) $U_R$ range 4 to 7: While maintaining large-amplitude vibrations, 
gradual amplitude attenuation emerges with increasing $U_R$. 
4) $U_R$ transition 7 to 8: The vortex shedding frequency substantially 
exceeds the natural frequency, effectively suppressing resonance 
and restoring small-amplitude cyclic motion. 
This behavior fundamentally demonstrates the lock-in phenomenon 
characteristic of VIV systems, 
where maximum vibration amplitudes occur when the Strouhal frequency 
synchronizes with the structural frequency. 

While the primary objective of this study focuses on validating the accuracy 
of the proposed moving mesh FSI approach, 
rather than exhaustively investigating the underlying mechanisms of SJs-based 
VIV suppression, 
we strategically select a resonant case 
$\mathrm{Re} = 150, m^* = 2, c^* = 0. 0, U_R = 4. 0 $ 
to evaluate the suppression efficacy. 
In the case of SJs -based VIV suppression , 
the two most common control parameters are SJs velocity magnitude 
and SJs actuation frequency. 
Notably, it is obvious that increasing the SJs velocity 
can impact the boundary layer more strongly and thus suppress 
the VIV better, 
but the variation of the SJs frequency affects the 
suppression effect in a less obvious way. 
This motivates our investigation into the intrinsic 
relationship between SJs actuation frequency and 
suppression performance, 
providing critical insights for active flow control optimization. 
\begin{figure}
  \includegraphics[width=0.45\textwidth]{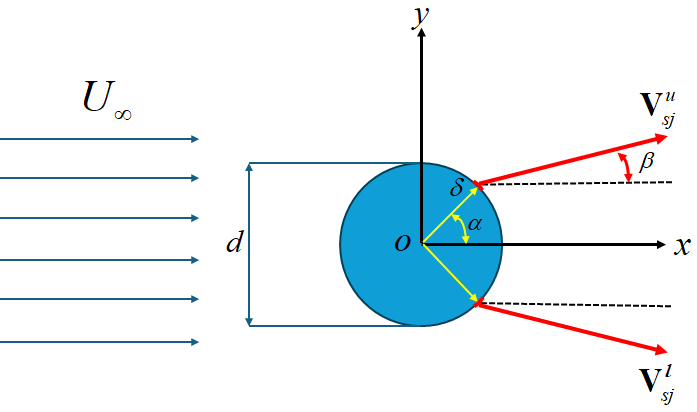}
  \caption{\label{fig:14} The sketch of Sjs-based VIV suppression. }
\end{figure}
The case of Sjs-based VIV suppression is shown in Fig.~\ref{fig:14}. 
The jet is positioned at the quarter-arc length on the leeward side 
of the cylinder, with the jet orifice fixed at an angle of $\alpha = 45^{\circ}$. 
The jet velocity is defined as:
\begin{equation}
  \begin{aligned}
    & \mathbf{V}_{sj}^{u}={{U}_{\max }}\sin (2\pi {{f}_{sj}}\left( t-{{t}_{sj}} \right)+{{\phi }_{u}}){{e}^{i\beta }} , \\ 
    & \mathbf{V}_{sj}^{l}={{U}_{\max }}\sin (2\pi {{f}_{sj}}\left( t-{{t}_{sj}} \right)+{{\phi }_{l}}){{e}^{i(-\beta )}}  . 
  \end{aligned}
  \label{eq:30}
\end{equation}

The superscript or subscripts \( u \) and \( l \) denote the upper jet and lower jet, 
respectively. Here, \( U_{\max} \) and \( f_{sj} \) represent 
the maximum jet velocity and jet frequency, \( t_{sj} \) 
specifies the activation time of the jet, and \( \phi_u \), \( \phi_l \) indicate 
the phase angles at jet initiation. 
In this study, these phases are set to \( \phi_u = \phi_l = 0 \). 
Additionally, the jet direction is fixed at \( \beta = 0^\circ \), 
corresponding to the horizontal direction. 
The dimensionless actuation frequency and momentum coefficient 
of the SJs can be expressed as:
\begin{equation}
  {{f}^{*}}=\frac{{{f}_{sj}}}{{{f}_{v}}}, {{C}_{\mu }}=\frac{2{{U}_{\max }}^{2}\delta }{{{u}_{\infty }}^{2}d}, 
  \label{eq:31}
\end{equation}
where the length of jet orifice is $\delta$, in this study $\delta =\frac{\pi d}{32}$. 

\begin{figure*}
  \begin{minipage}[b]{0.45\textwidth}
    \centering
    \includegraphics[height=7cm]{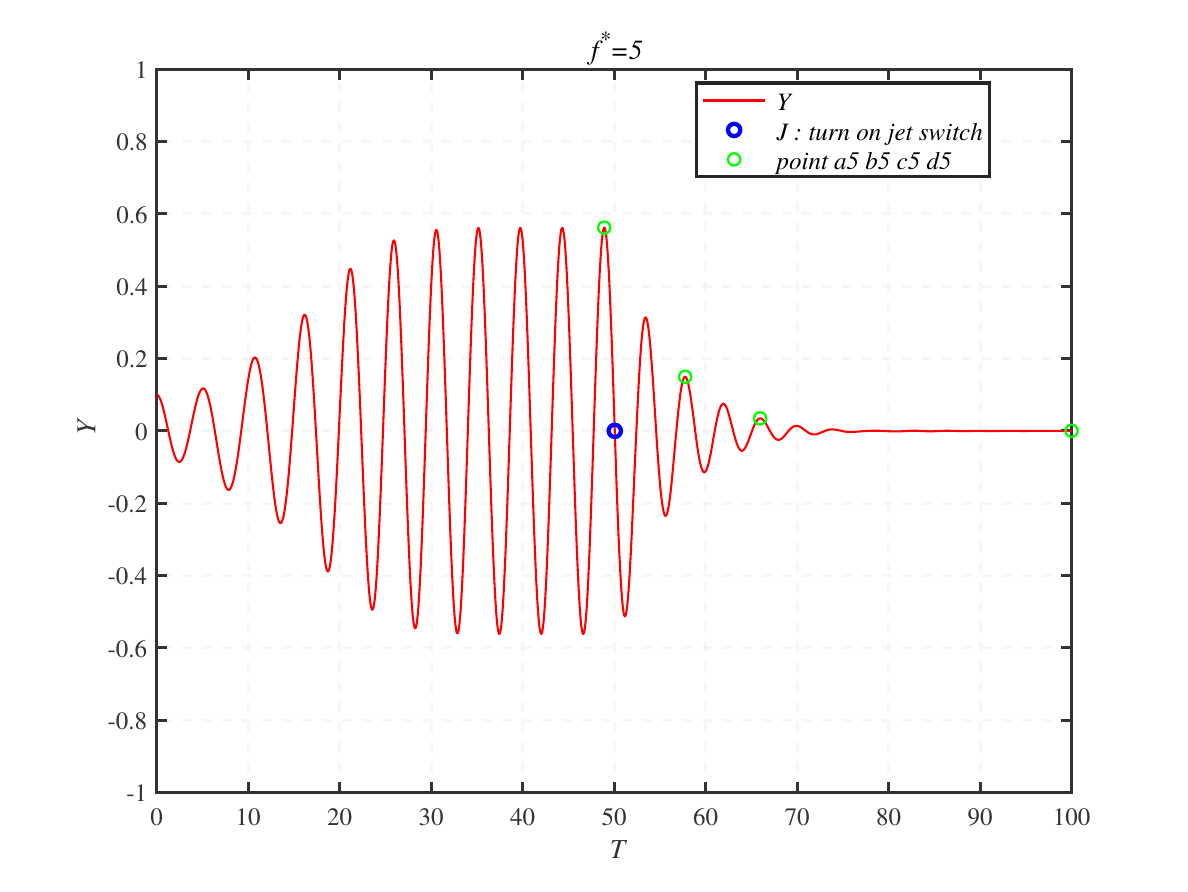} 
    \caption*{(a) $f^{*}=5$}
  \end{minipage}
  \hfill 
  \begin{minipage}[b]{0.45\textwidth}
    \centering
    \includegraphics[height=7cm]{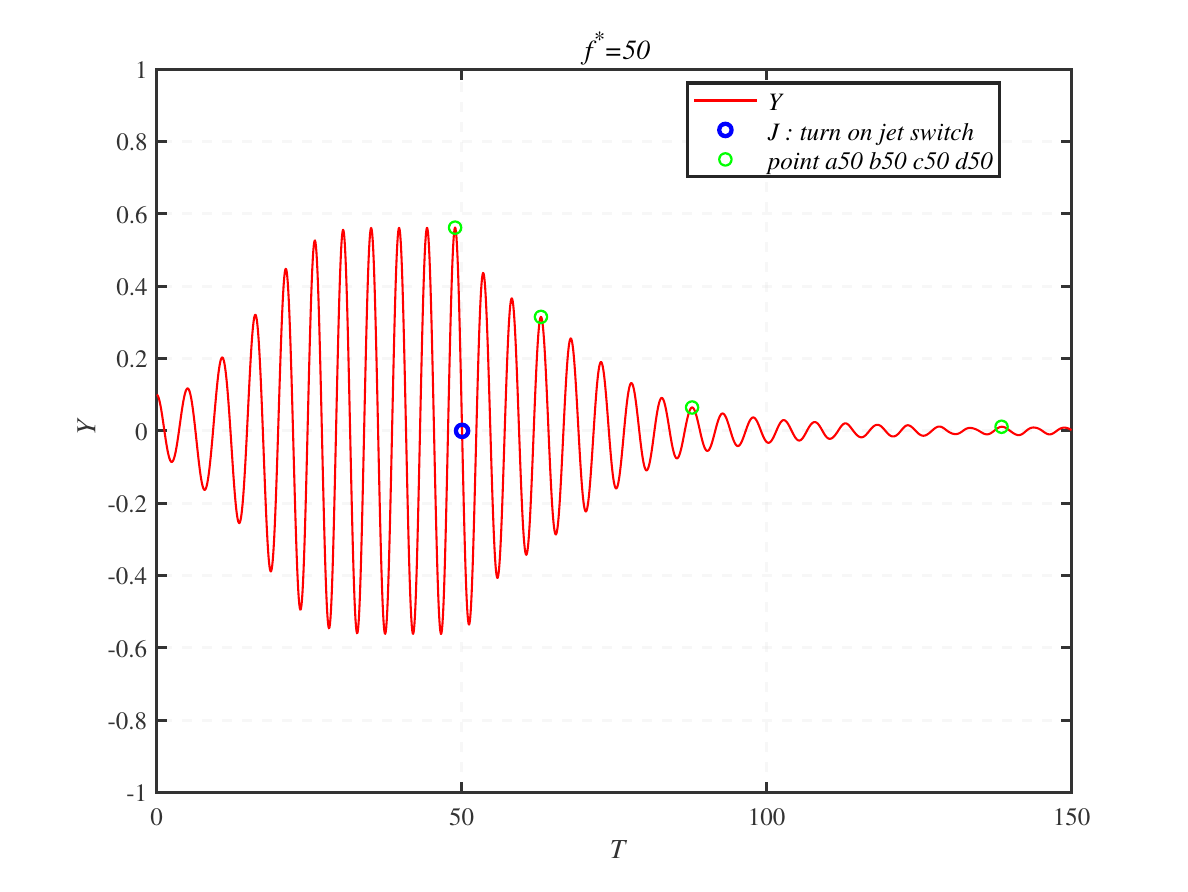} 
    \caption*{(b) $f^{*}=50$}
  \end{minipage}
  \begin{minipage}[b]{0.45\textwidth}
    \centering
    \includegraphics[height=7cm]{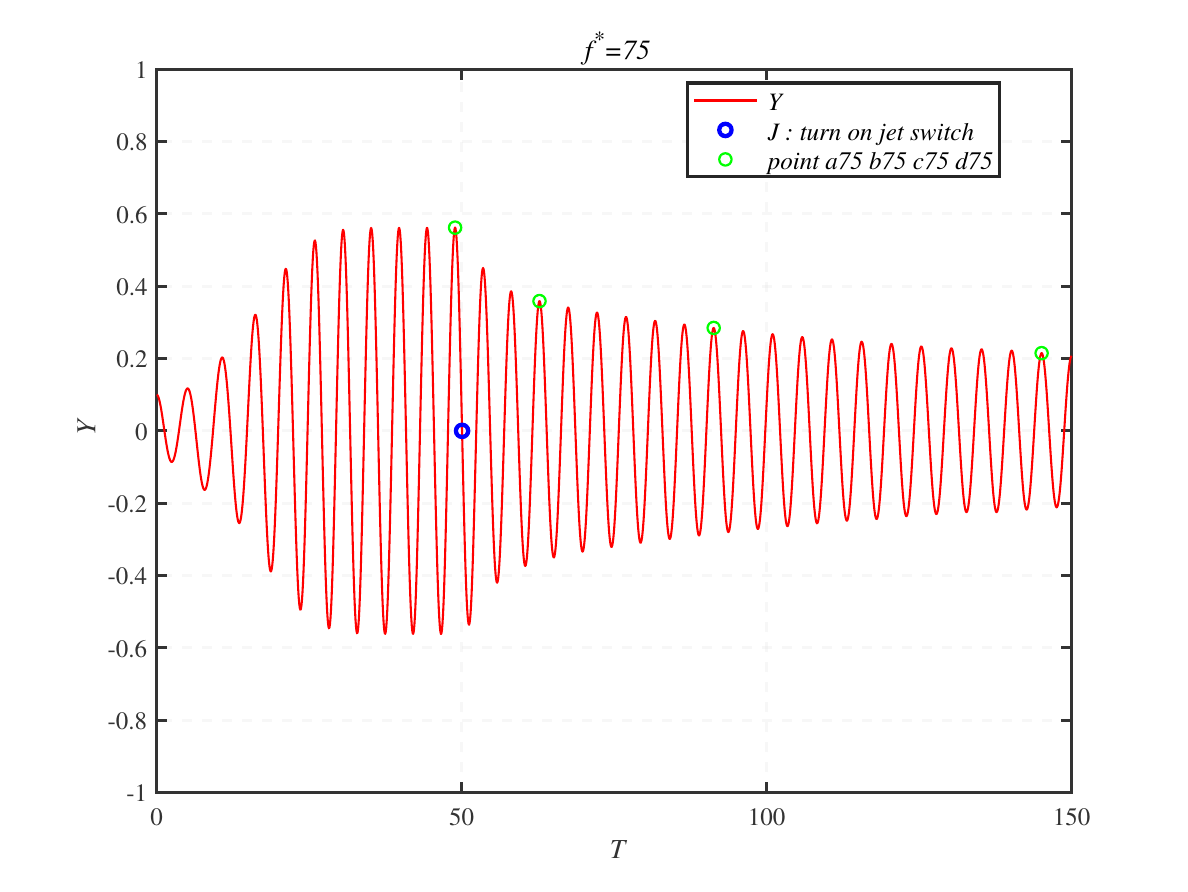} 
    \caption*{(c) $f^{*}=75$}
  \end{minipage}
  \hfill 
  \begin{minipage}[b]{0.45\textwidth}
    \centering
    \includegraphics[height=7cm]{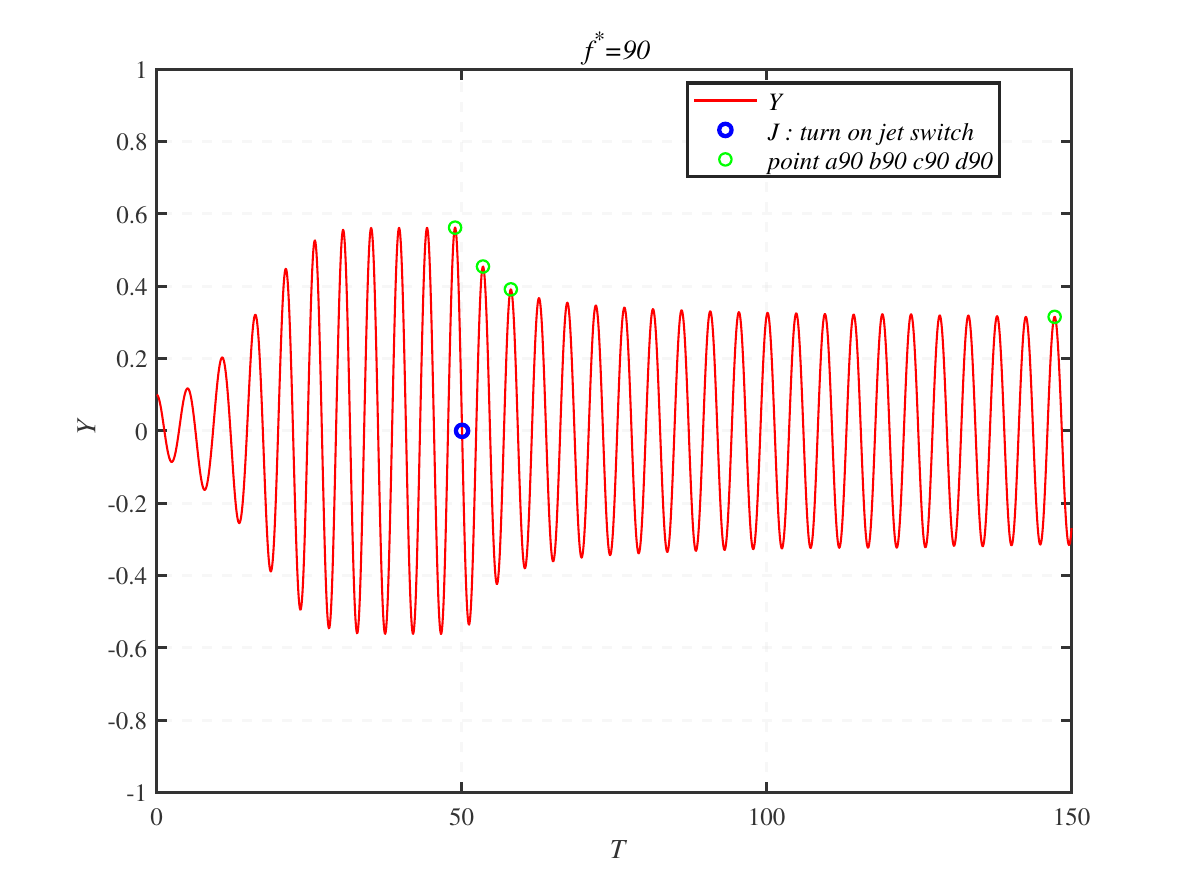} 
    \caption*{(d) $f^{*}=90$}
  \end{minipage}
  \caption{\label{fig15} The time evolution of transverse displacement of the cylinder with $f^{*}=5, 50, 75, 90$. }
\end{figure*}

Representative cases with dimensionless actuation frequency 
$f^{*}=5, 50, 75, 90$ and momentum coefficient $C_{\mu}=3. 0$ are 
investigated to demonstrate the SJs-based VIV suppression 
efficacy while exploring the frequency-dependent suppression characteristics. 
To accelerate the cylinder's transition to a steady resonant state, 
the initial dimensionless displacement is set as \( Y(T) = Y(0) = 0. 1 \). 
After the VIV reaches a steady state, 
the jet actuator is activated when the cylinder's displacement transitions 
from positive to zero. Specifically, 
the jet is triggered at the dimensionless time:
$T_{\text{sj}} = \frac{t_{\text{sj}}}{d / u_{\infty}} = 50. 0698$
where \( t_{\text{sj}} \) is the physical jet activation time, 
\( d \) is the cylinder diameter, 
and \( u_{\infty} \) is the free-stream velocity. 
The time evolution of transverse displacement of the cylinder is shown 
in Fig.~\ref{fig15}. The coordinates of points \( a \), \( b \), \( c \), and \( d \) 
for each case in Fig.~\ref{fig15} are listed in Table. ~\ref{tab:3}.

\begin{table*}
  \centering
  \caption{The coordinates of points \( a \), \( b \), \( c \), and \( d \). }
  \begin{ruledtabular}
  \begin{tabular}{c|cccc}
    &a5&b5&c5&d5\\
    &(48. 9236, 0. 562)&	(57. 7611, 0. 1497)	&(65. 9610, 0. 0346)	&(100, 0. 0001)\\
    &a50	&b50	&c50&	d50\\
    &(48. 9236, 0. 562)&	(63. 0031, 0. 3152)&	(87. 7822, 0. 0646)	&(138. 5217, 0. 0113)\\
   (T, Y) &a75&	b75&	c75	&d75\\
    &(48. 9236, 0. 562)	&(62. 7653, 0. 3589)	&(91. 3238, 0. 2846)	&(145. 0911, 0. 2149)\\
    &a90	&b90&	c90&	d90\\
    &(48. 9236, 0. 562)	&(53. 4984, 0. 4545)	&(58. 0690, 0. 3914)&	(147. 2236, 0. 3153)\\
  \end{tabular}
  \end{ruledtabular}
  \label{tab:3}
\end{table*}

\newpage
\newpage

\begin{figure}
  \begin{minipage}[b]{0.4\textwidth}
    \centering
    \includegraphics[height=6cm]{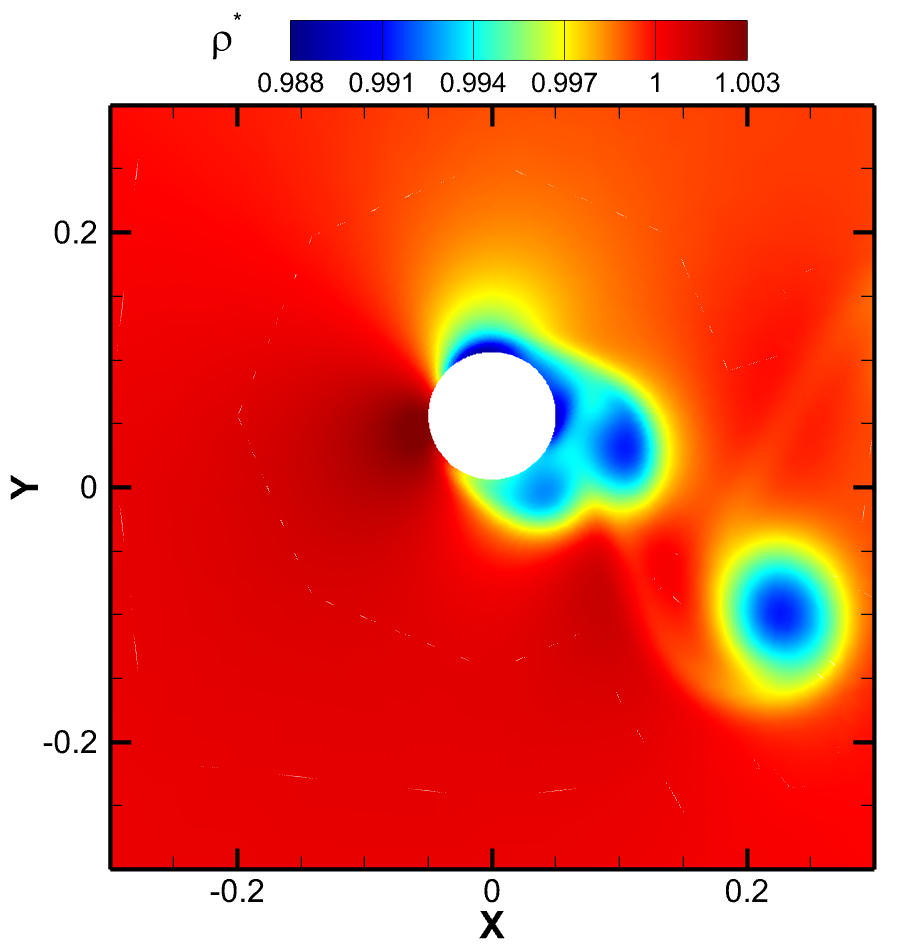} 
    \caption*{(a) $\rho^*$ contour}
  \end{minipage}
  \begin{minipage}[b]{0.45\textwidth}
    \centering
    \includegraphics[height=5.5cm]{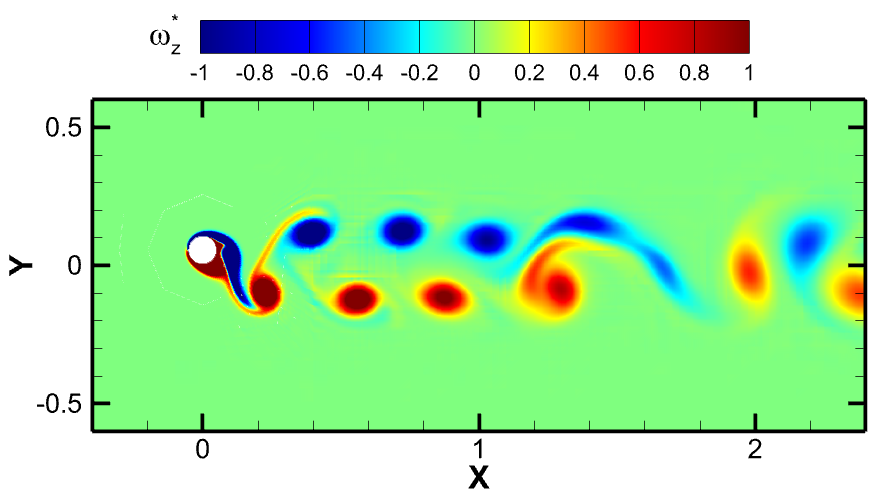} 
    \caption*{(b) $\omega_z^*$ contour}
  \end{minipage}
  \caption{\label{fig16} $\omega_z^*$ and $\rho^*$ contour plots at point $a$}
\end{figure}

Prior to SJs activation, 
all four cases exhibit identical flowfield evolution patterns. 
The SJs are synchronously activated at non-dimensional time $T$=50. 0698, 
marked as point J. 
Post-activation, divergent flowfield developments emerge across cases 
due to their distinct dimensionless actuation frequencies. 
As evidenced by the data, 
cases $a5$, $a50$, $a75$, and $a90$ share identical pre-activation peak 
displacement coordinates (T, Y) = (48. 9236, 0. 562). 
These cases maintain completely congruent flowfield configurations 
at this critical instant, hereafter collectively designated 
as reference state "$a$". At this transitional phase, 
two coherent vortex streets are observed downstream of the cylinder. 
Fig.~\ref{fig16} correspondingly presents the non-dimensional vorticity $\omega_z^*$ 
and non-dimensional density $\rho^*$(where $\rho^*=\frac{\rho}{\rho_{\infty}}$) contour plots characterizing state "$a$".

\begin{figure*}
  \begin{minipage}[b]{0.3\textwidth}
    \centering
    \includegraphics[height=6cm]{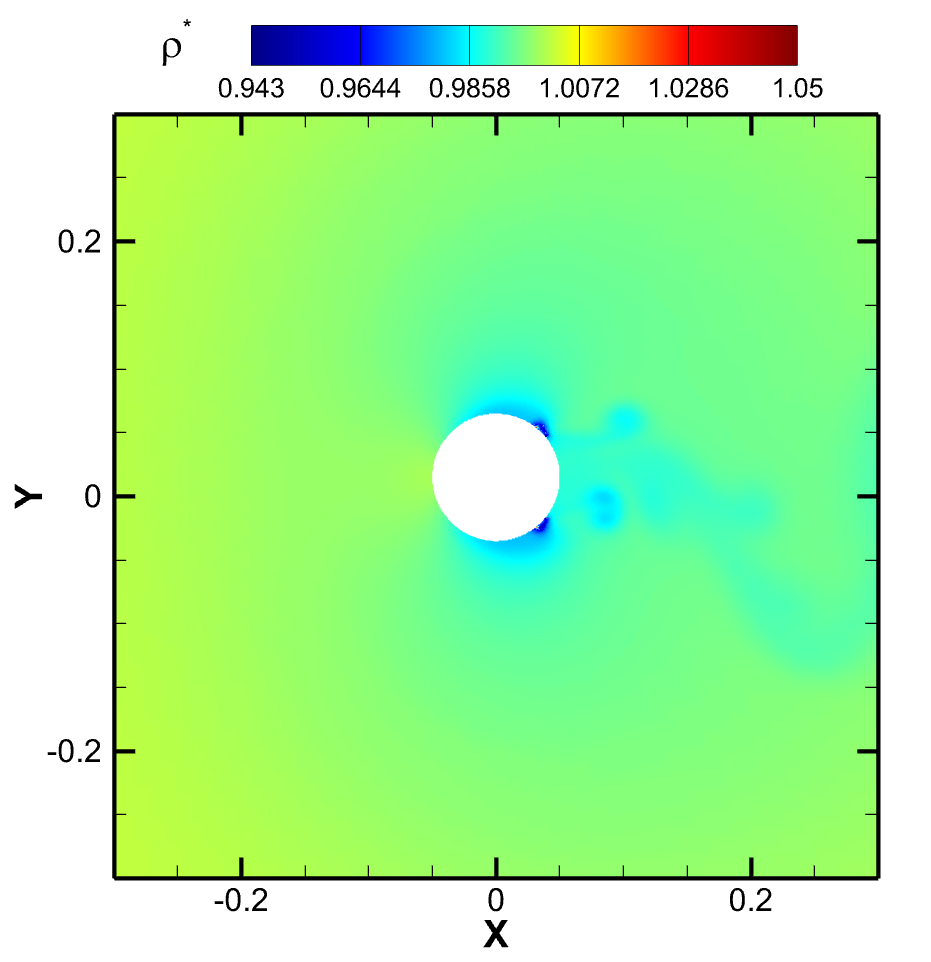} 
    \caption*{(a) $\rho^*$ contour at points $b5$}
  \end{minipage}
  \hfill 
  \begin{minipage}[b]{0.6\textwidth}
    \centering
    \includegraphics[height=6cm]{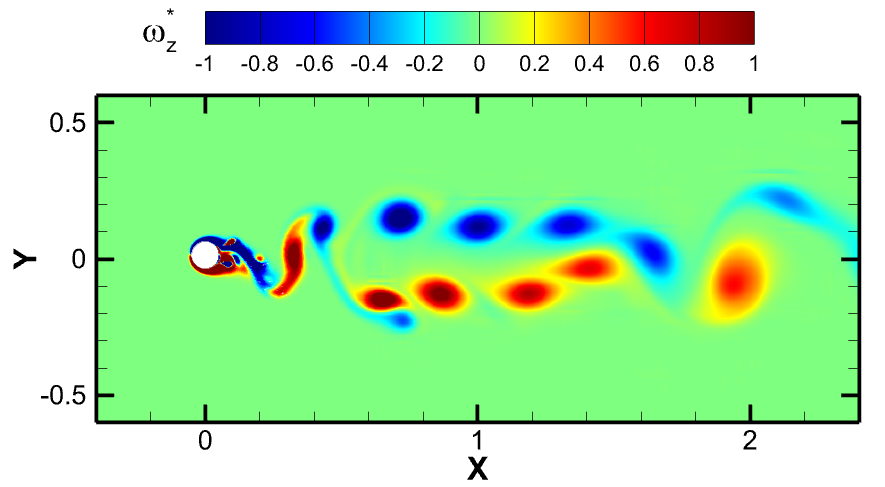} 
    \caption*{(b) $\omega_z^*$ contour at points $b5$}
  \end{minipage}
\begin{minipage}[b]{0.3\textwidth}
  \centering
  \includegraphics[height=6cm]{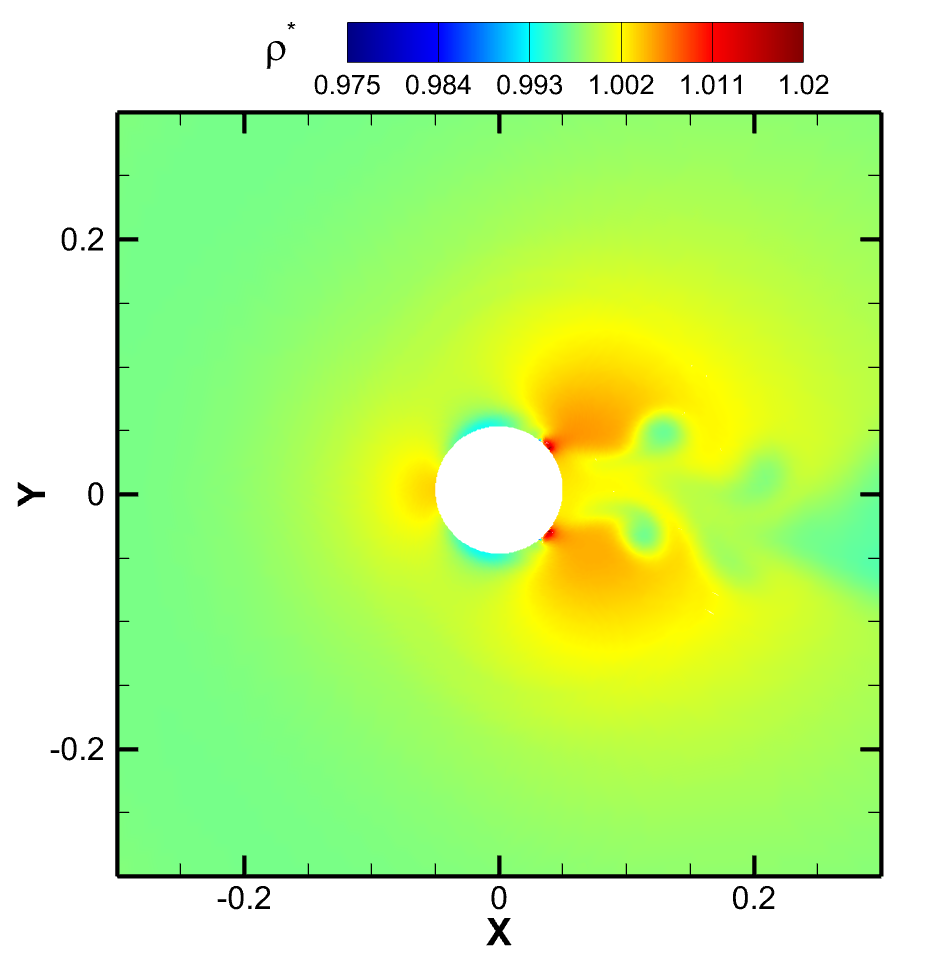} 
  \caption*{(c) $\rho^*$ contour at points $c5$}
\end{minipage}
\hfill 
\begin{minipage}[b]{0.6\textwidth}
  \centering
  \includegraphics[height=6cm]{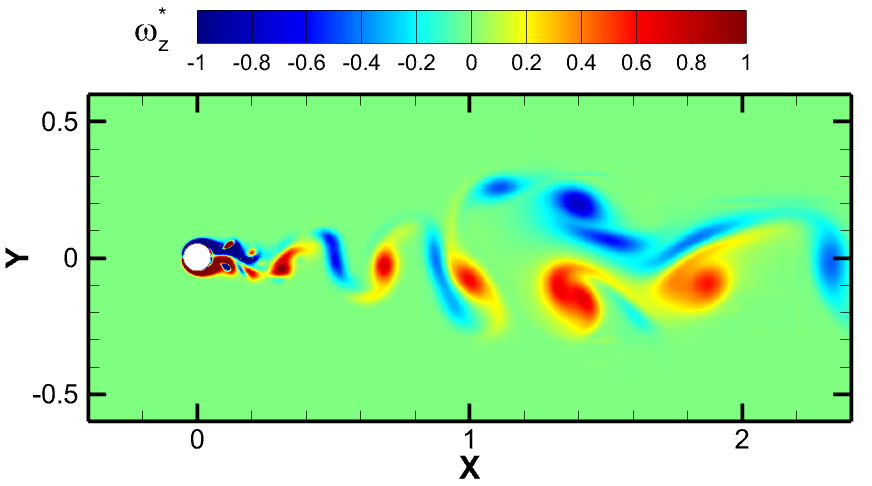} 
  \caption*{(d) $\omega_z^*$ contour at points $c5$}
\end{minipage}
\begin{minipage}[b]{0.3\textwidth}
  \centering
  \includegraphics[height=6cm]{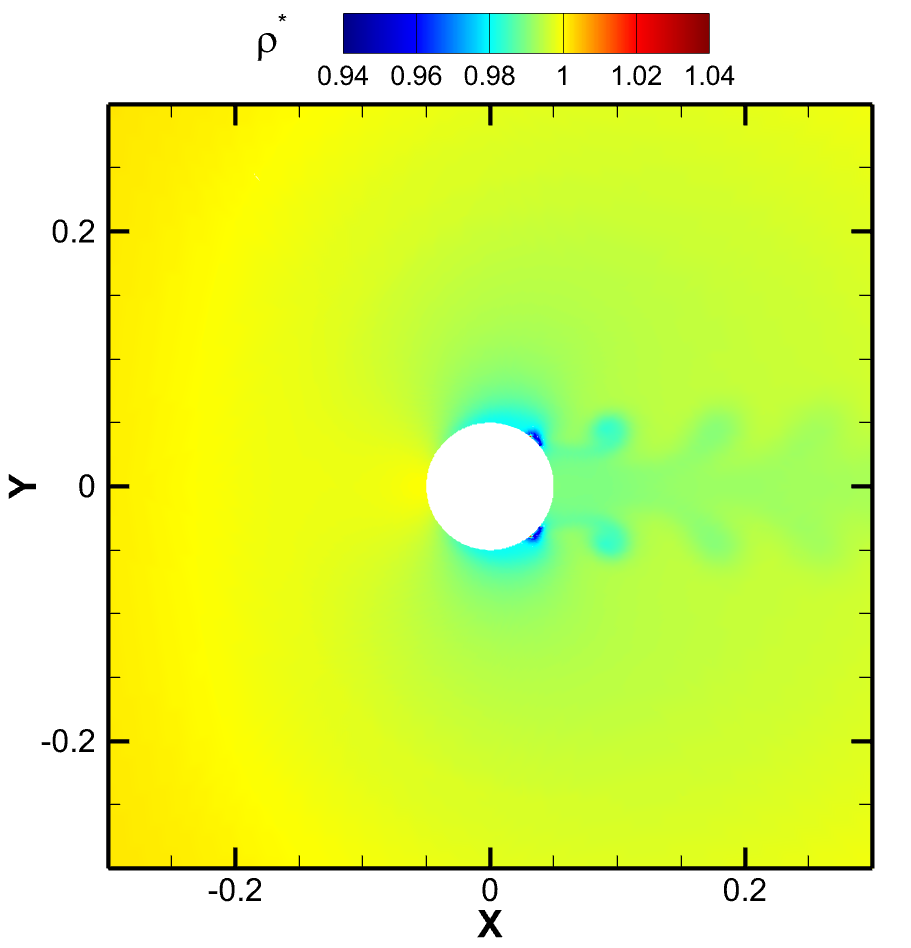} 
  \caption*{(e) $\rho^*$ contour at points $d5$}
\end{minipage}
\hfill 
\begin{minipage}[b]{0.6\textwidth}
  \centering
  \includegraphics[height=6cm]{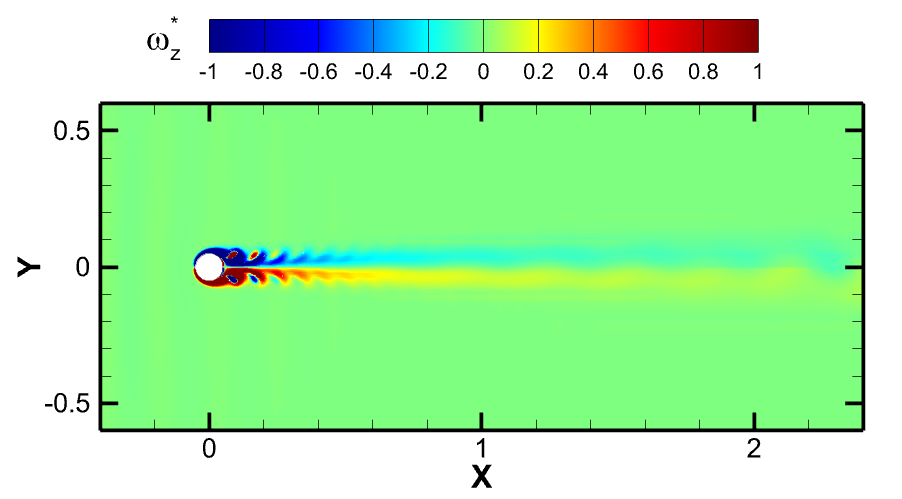} 
  \caption*{(f) $\omega_z^*$ contour at points $d5$}
\end{minipage}
  \caption{\label{fig17} $\omega_z^*$ and $\rho^*$ contour plots at points $b5$, $c5$, and $d5$}
\end{figure*}

Under the case of $f^*$=5, Fig.~\ref{fig16}(a) demonstrates that SJs induces 
rapid attenuation in the dimensionless displacement amplitude Y of the cylinder with 
increasing dimensionless time T. 
Within two oscillation cycles after passing point $a5$, the cylinder reaches point $b5$ 
at $T$=57. 7611, exhibiting a significantly reduced displacement amplitude $Y_{b5}=0. 1497$. 
This amplitude represents approximately one-fourth of the $Y_a$, 
attributed to SJs-induced boundary layer impact which elongates the 
shedding vortices and initiates their fragmentation. 
Subsequent evolution reveals further amplitude reduction to $Y_{c5}$=0. 0346 at point $c5$
 ($T$=65. 9610), 
 accompanied by progressively stretched and disintegrated wake vortices. 
 When $T$ exceeds 80 (point $d5$), the $Y$ stabilizes near zero with minimal fluctuations, 
 indicating complete VIV suppression through SJs control. 
 This stabilization corresponds to the establishment of a symmetric flow field maintained 
 by sustained SJs momentum injection. 
 The corresponding vortex dynamics evolution is quantitatively illustrated through 
 dimensionless vorticity $\omega_z^*$ and dimensionless density $\rho^*$ 
 contours at characteristic points $b5$, $c5$, and $d5$ in Fig.~\ref{fig17}. 

\begin{figure*}
  \begin{minipage}[b]{0.3\textwidth}
    \centering
    \includegraphics[height=6cm]{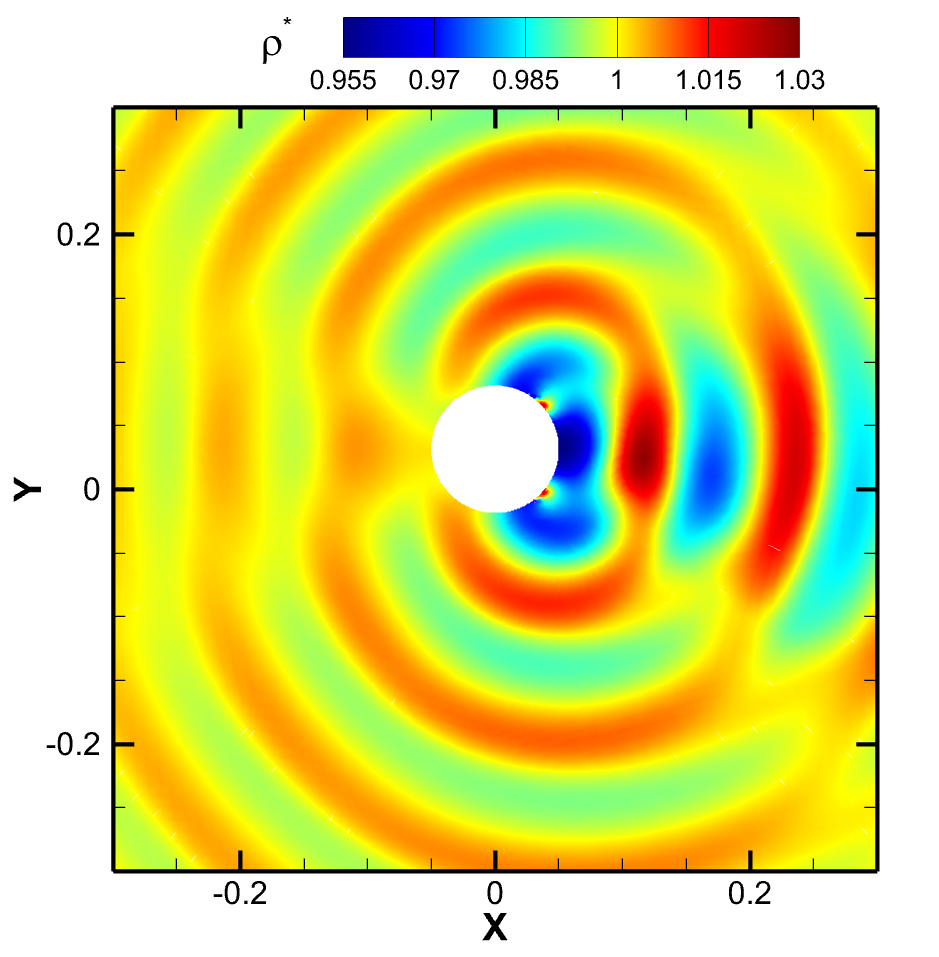} 
    \caption*{(a) $\rho^*$ contour at points $b50$}
  \end{minipage}
  \hfill 
  \begin{minipage}[b]{0.6\textwidth}
    \centering
    \includegraphics[height=6cm]{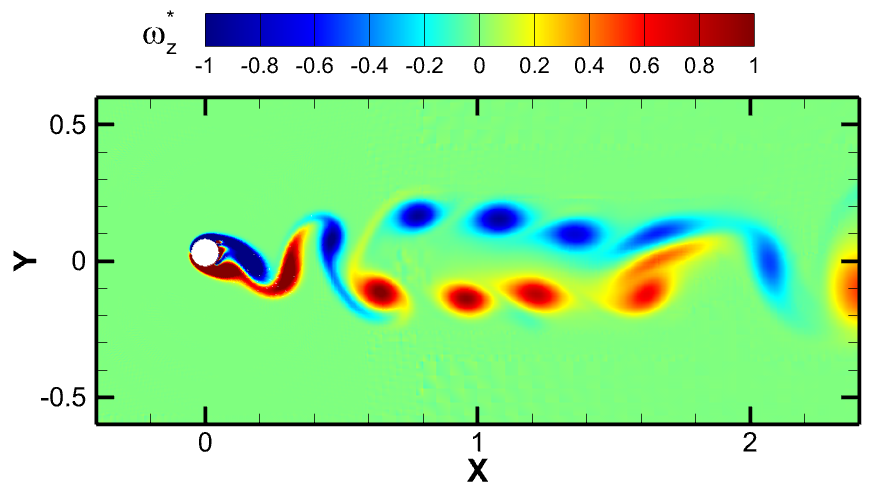} 
    \caption*{(b) $\omega_z^*$ contour at points $b50$}
  \end{minipage}
\begin{minipage}[b]{0.3\textwidth}
  \centering
  \includegraphics[height=6cm]{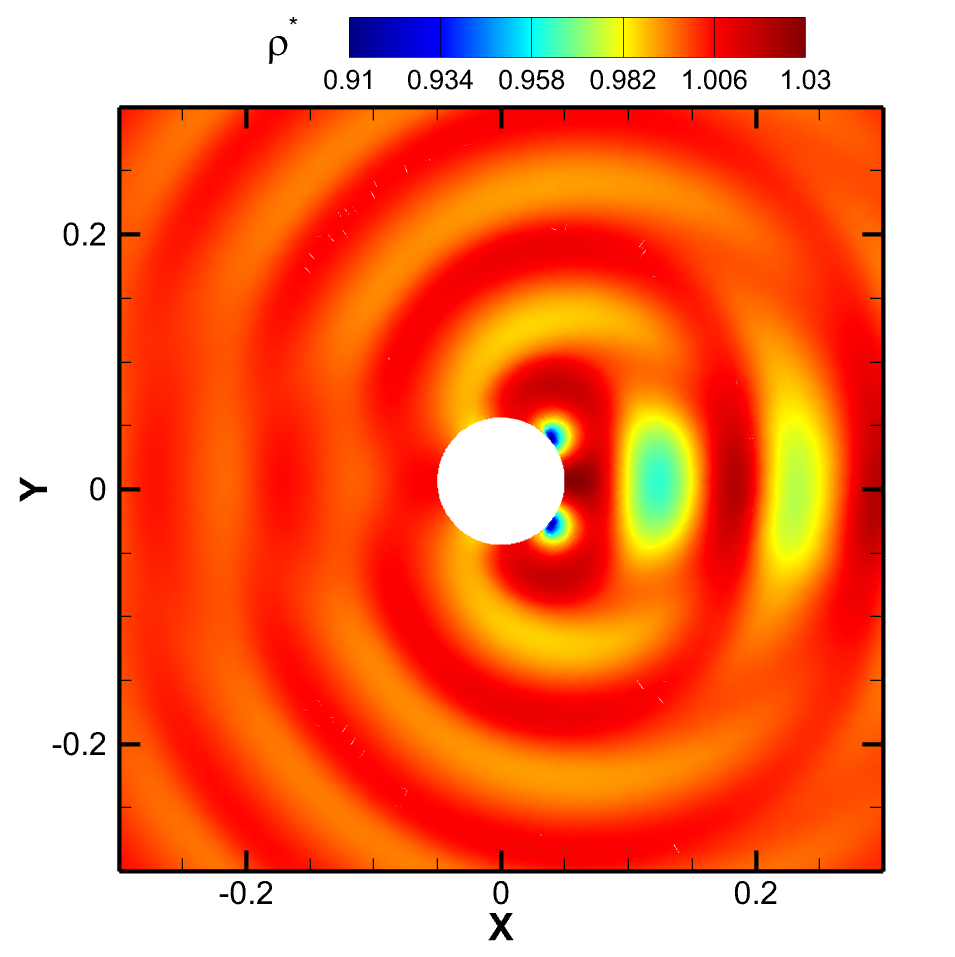} 
  \caption*{(c) $\rho^*$ contour at points $c50$}
\end{minipage}
\hfill 
\begin{minipage}[b]{0.6\textwidth}
  \centering
  \includegraphics[height=6cm]{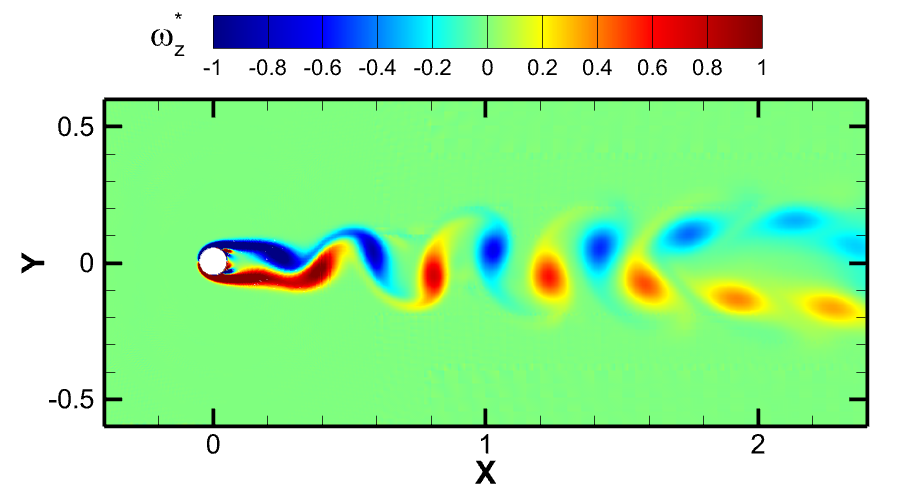} 
  \caption*{(d) $\omega_z^*$ contour at points $c50$}
\end{minipage}
\begin{minipage}[b]{0.3\textwidth}
  \centering
  \includegraphics[height=6cm]{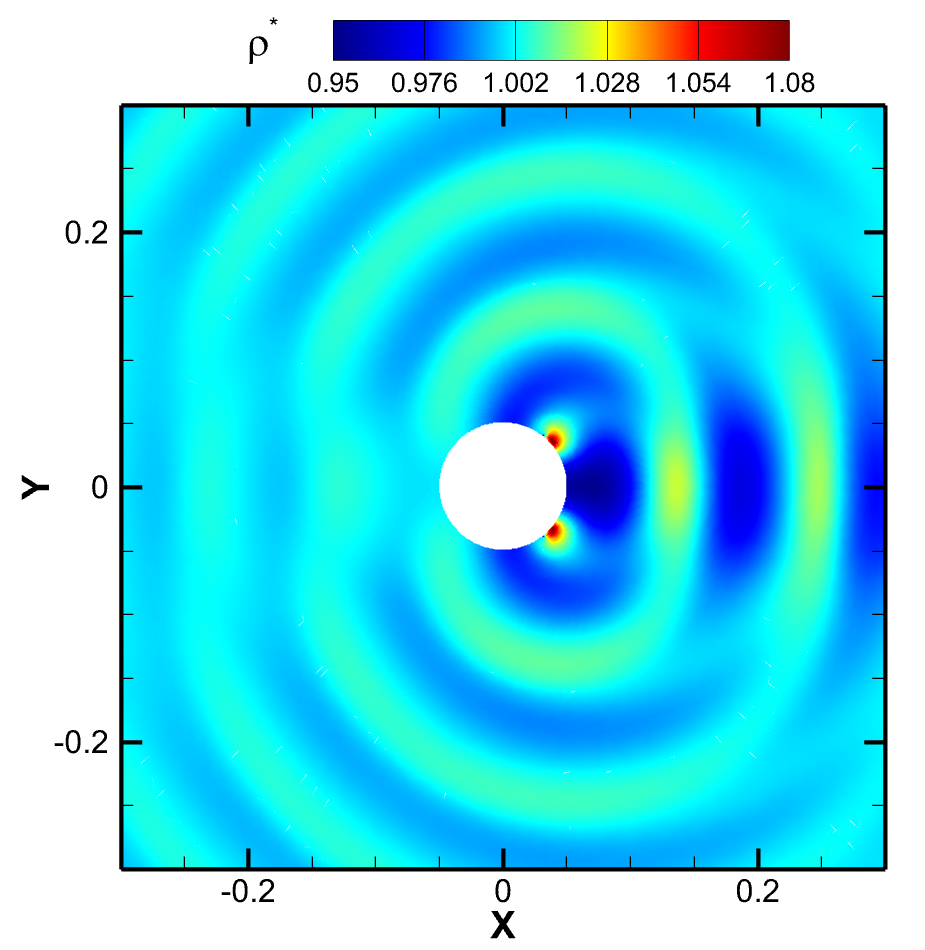} 
  \caption*{(e) $\rho^*$ contour at points $d50$}
\end{minipage}
\hfill 
\begin{minipage}[b]{0.6\textwidth}
  \centering
  \includegraphics[height=6cm]{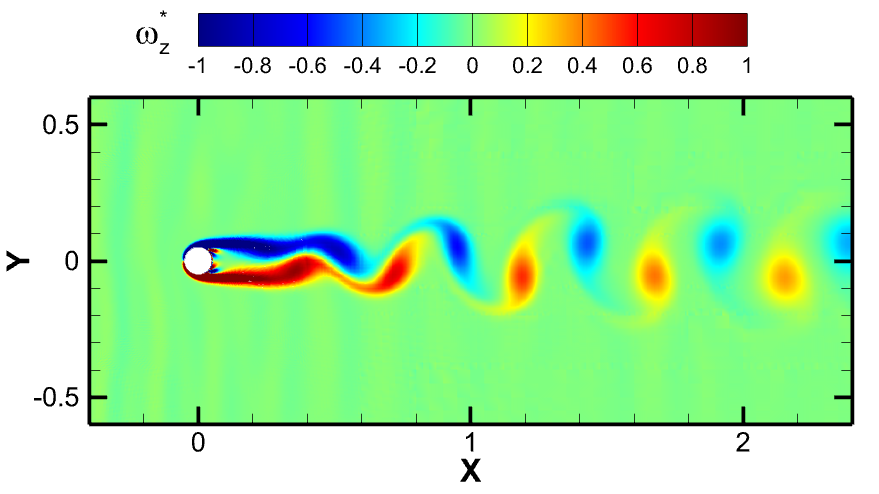} 
  \caption*{(f) $\omega_z^*$ contour at points $d50$}
\end{minipage}
  \caption{\label{fig18} $\omega_z^*$ and $\rho^*$ contour plots at points $b50$, $c50$, and $d50$}
\end{figure*}

Under the case of $f^*$=50, Fig.~\ref{fig15}(b) reveals that SJs 
induces marked attenuation in the dimensionless displacement amplitude $Y$ of the cylinder 
with increasing dimensionless time $T$, though the attenuation rate is comparatively slower 
than observed at $f^*$=5. At characteristic point $d50$ ($Y_{d50}$=0. 0113), 
the VIV is effectively suppressed, achieving near-complete stabilization. 
The evolutionary characteristics of flow structures are 
systematically documented through dimensionless vorticity $\omega_z^*$ and 
dimensionless density $\rho^*$ contours at characteristic 
point $b50$, $c50$, and $d50$ in Fig.~\ref{fig18}. 
The vorticity dynamics captured in Fig.~\ref{fig18} demonstrate 
elongated attached vortices along cylinder two sides that maintain structural integrity 
without fragmentation tendencies. Corresponding density field analysis identifies 
a dual energy transfer mechanism: 
partial SJs energy propagates as acoustic waves 
while the remainder directly impacts boundary layer modulation. 

\begin{figure*}
  \begin{minipage}[b]{0.3\textwidth}
    \centering
    \includegraphics[height=6cm]{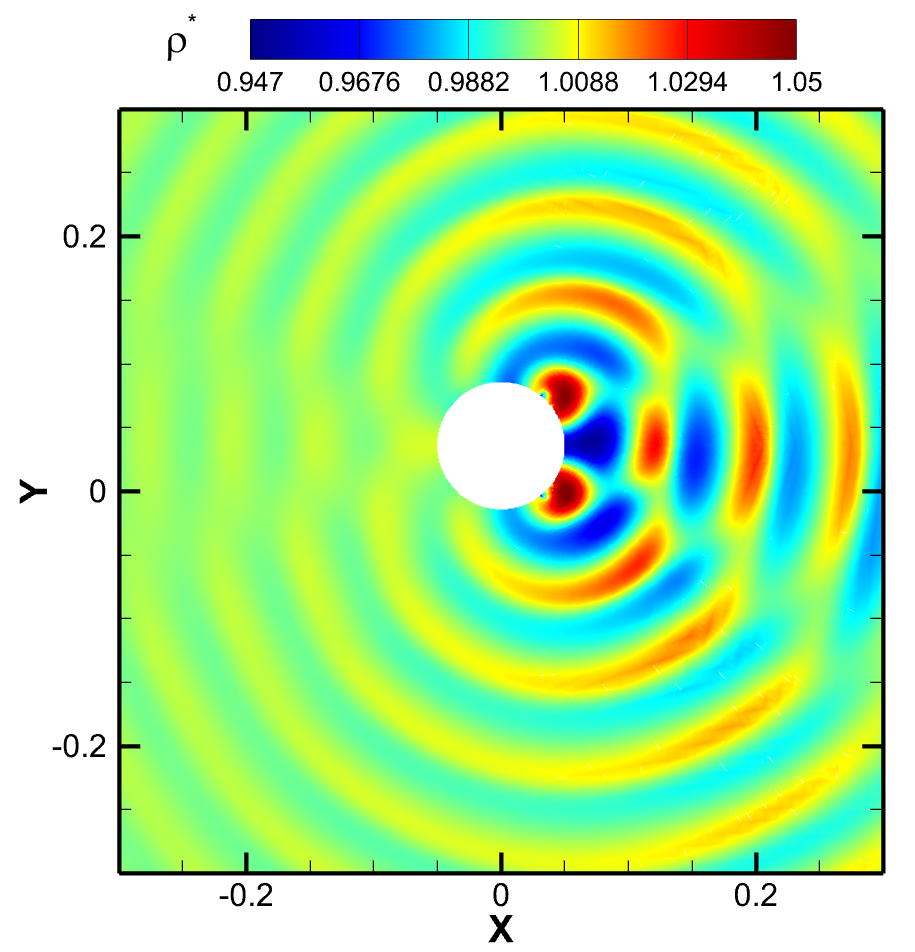} 
    \caption*{(a) $\rho^*$ contour at points $b75$}
  \end{minipage}
  \hfill 
  \begin{minipage}[b]{0.6\textwidth}
    \centering
    \includegraphics[height=6cm]{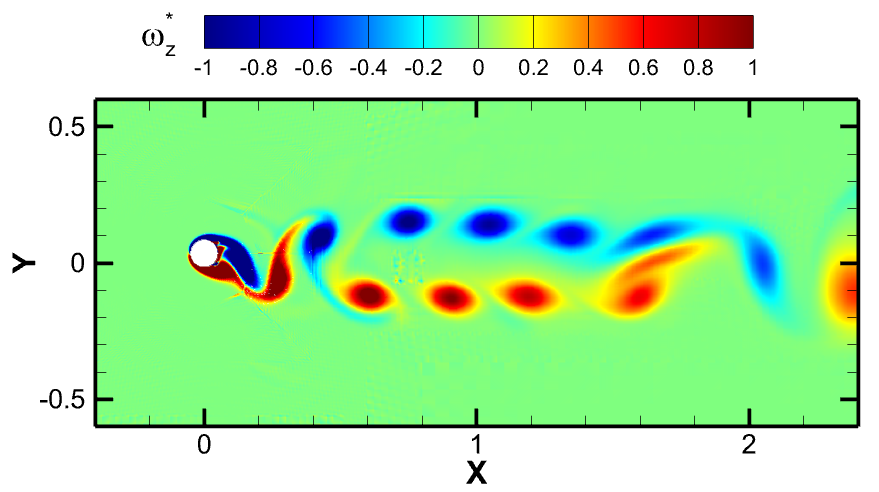} 
    \caption*{(b) $\omega_z^*$ contour at points $b75$}
  \end{minipage}
\begin{minipage}[b]{0.3\textwidth}
  \centering
  \includegraphics[height=6cm]{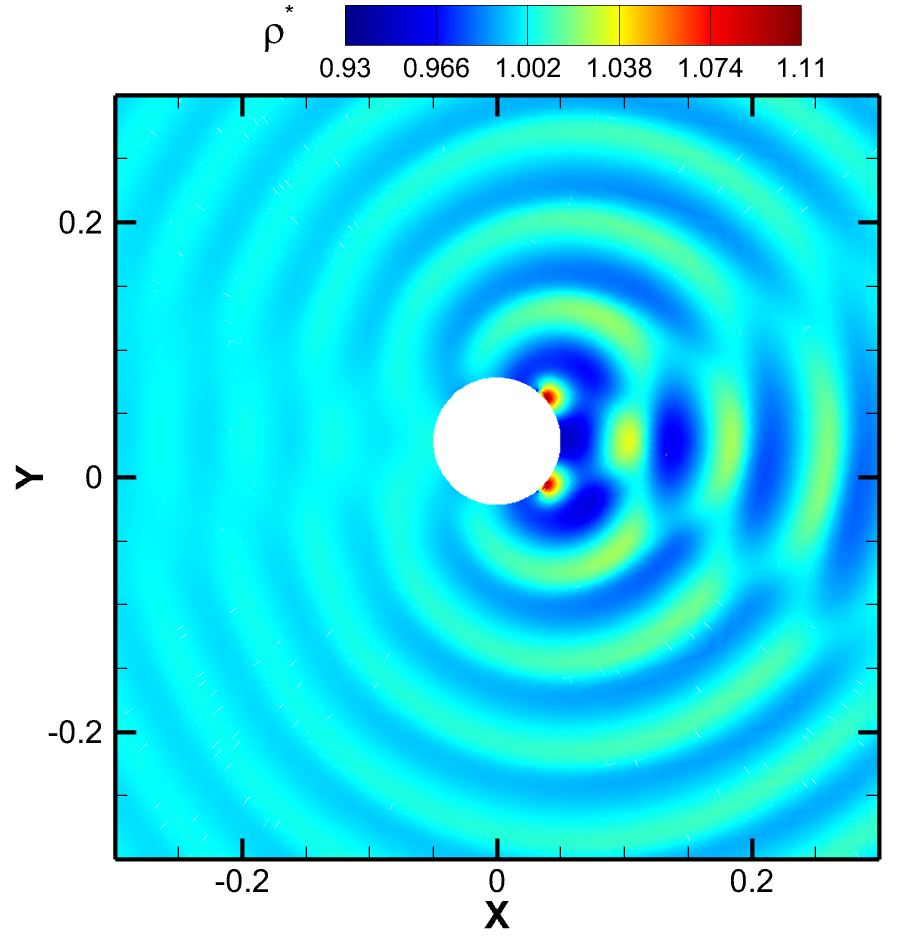} 
  \caption*{(c) $\rho^*$ contour at points $c75$}
\end{minipage}
\hfill 
\begin{minipage}[b]{0.6\textwidth}
  \centering
  \includegraphics[height=6cm]{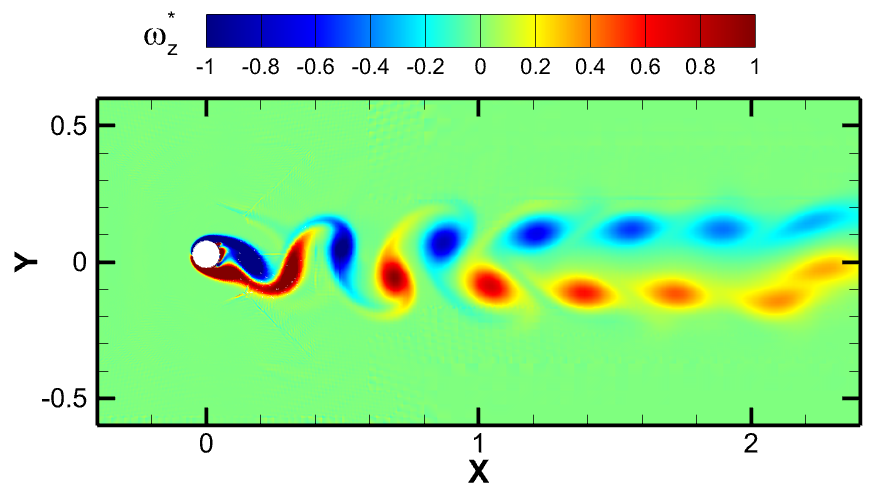} 
  \caption*{(d) $\omega_z^*$ contour at points $c75$}
\end{minipage}
\begin{minipage}[b]{0.3\textwidth}
  \centering
  \includegraphics[height=6cm]{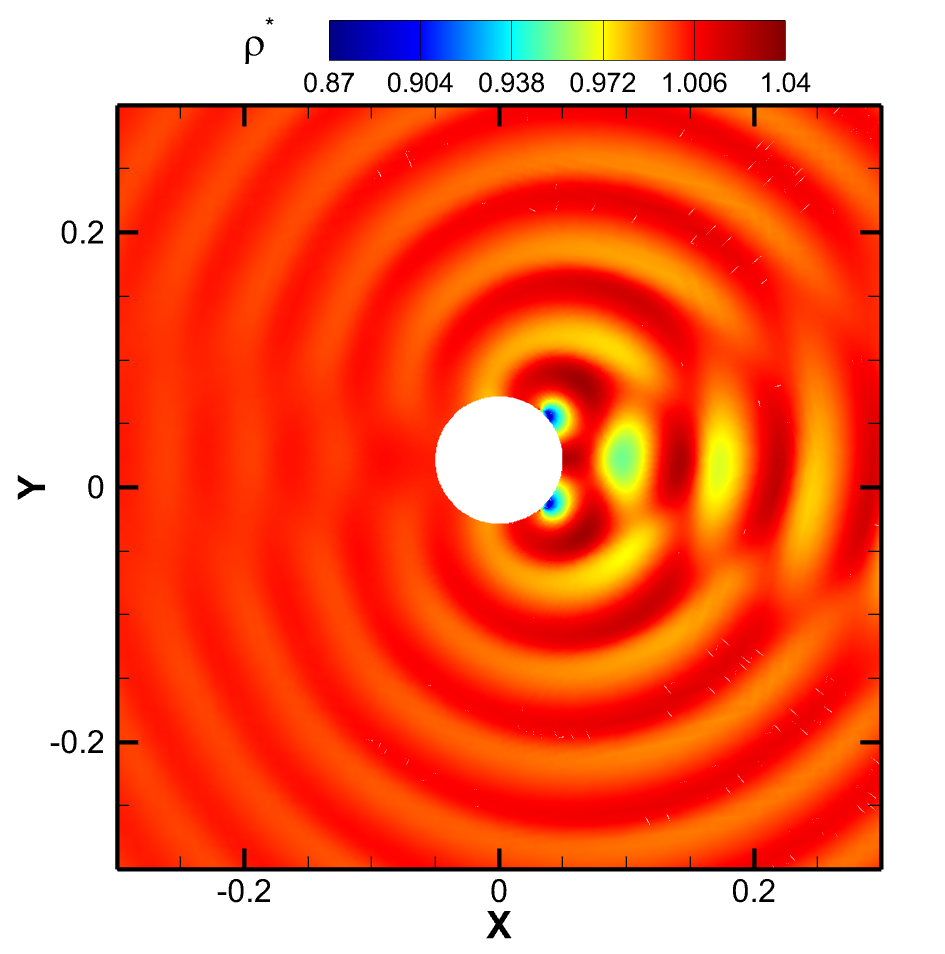} 
  \caption*{(e) $\rho^*$ contour at points $d75$}
\end{minipage}
\hfill 
\begin{minipage}[b]{0.6\textwidth}
  \centering
  \includegraphics[height=6cm]{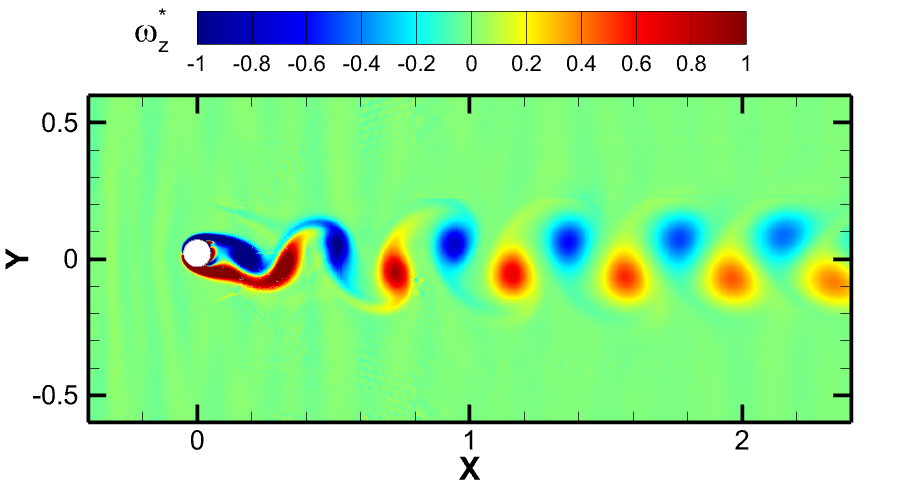} 
  \caption*{(f) $\omega_z^*$ contour at points $d75$}
\end{minipage}
  \caption{\label{fig19} $\omega_z^*$ and $\rho^*$ contour plots at points $b75$, $c75$, and $d75$}
\end{figure*}

Under the case $f^*$=75, Fig.~\ref{fig15}(c) demonstrates that following SJs activation, 
the amplitude of the cylinder's dimensionless displacement $Y$ exhibits slow attenuation 
with increasing dimensionless time $T$. At the point $d75$, 
the amplitude diminishes to $Y_{d75}=0. 2149$ with a decay rate approaching zero, 
indicating the SJs's insufficient capacity to completely suppress VIV. 
The evolutionary characteristics of flow structures are systematically 
documented through dimensionless vorticity $\omega_z^*$ and dimensionless 
density $\rho^*$ contours at characteristic point $b75$, $c75$, and $d75$ in Fig.~\ref{fig19}. 
The $\omega_z^*$ vorticity contours in Fig.~\ref{fig19} reveal marginally 
elongated vortices shedding from both sides of the cylinder, 
while two coherent vortex streets remain observable in the wake region. 
Comparative analysis of $\rho^*$ contours shows 
acoustic waves with notably shorter wavelengths than those observed in the $f^*$=50 case. 
This wavelength reduction enhanced energy's propagation through 
acoustic wave, where the majority of SJs energy radiates as acoustic waves 
while only a minor portion is utilized to impact the boundary layer for vibration suppression. 

\begin{figure*}
  \begin{minipage}[b]{0.3\textwidth}
    \centering
    \includegraphics[height=6cm]{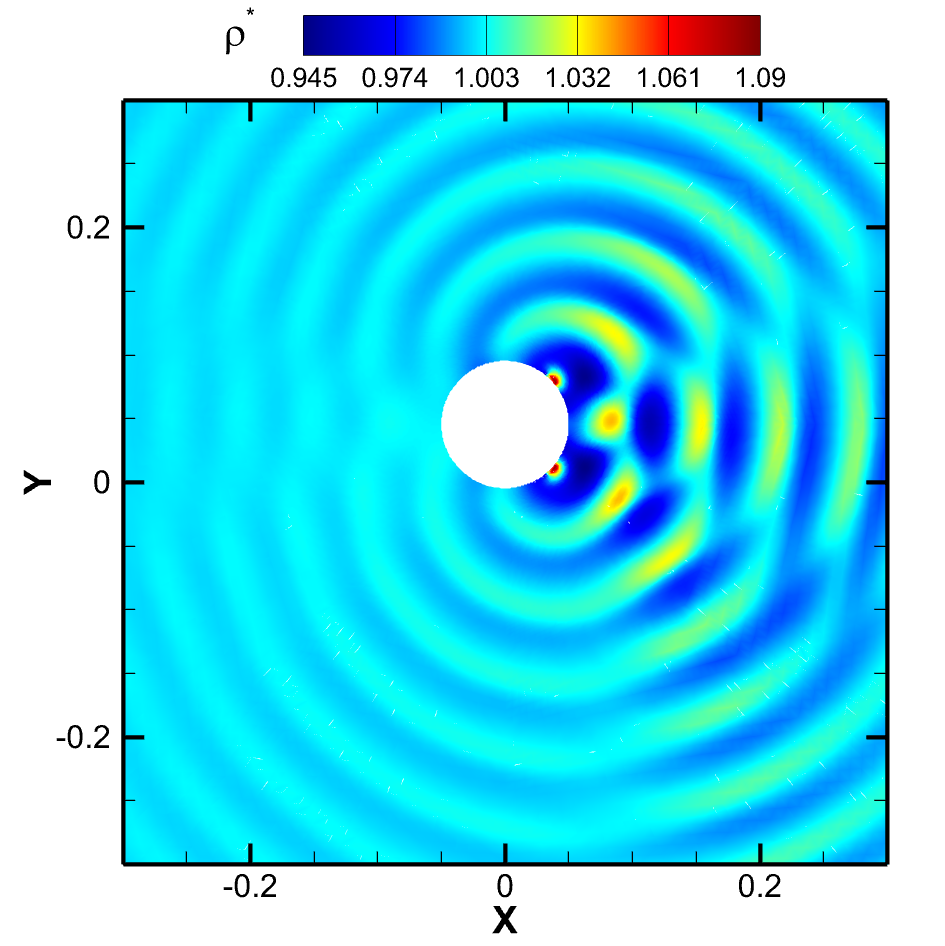} 
    \caption*{(a) $\rho^*$ contour at points $b90$}
  \end{minipage}
  \hfill 
  \begin{minipage}[b]{0.6\textwidth}
    \centering
    \includegraphics[height=6cm]{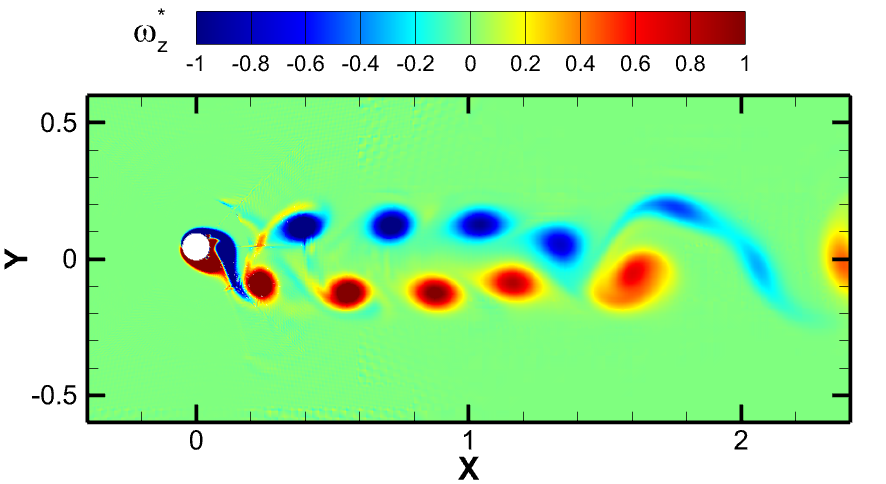} 
    \caption*{(b) $\omega_z^*$ contour at points $b90$}
  \end{minipage}
\begin{minipage}[b]{0.3\textwidth}
  \centering
  \includegraphics[height=6cm]{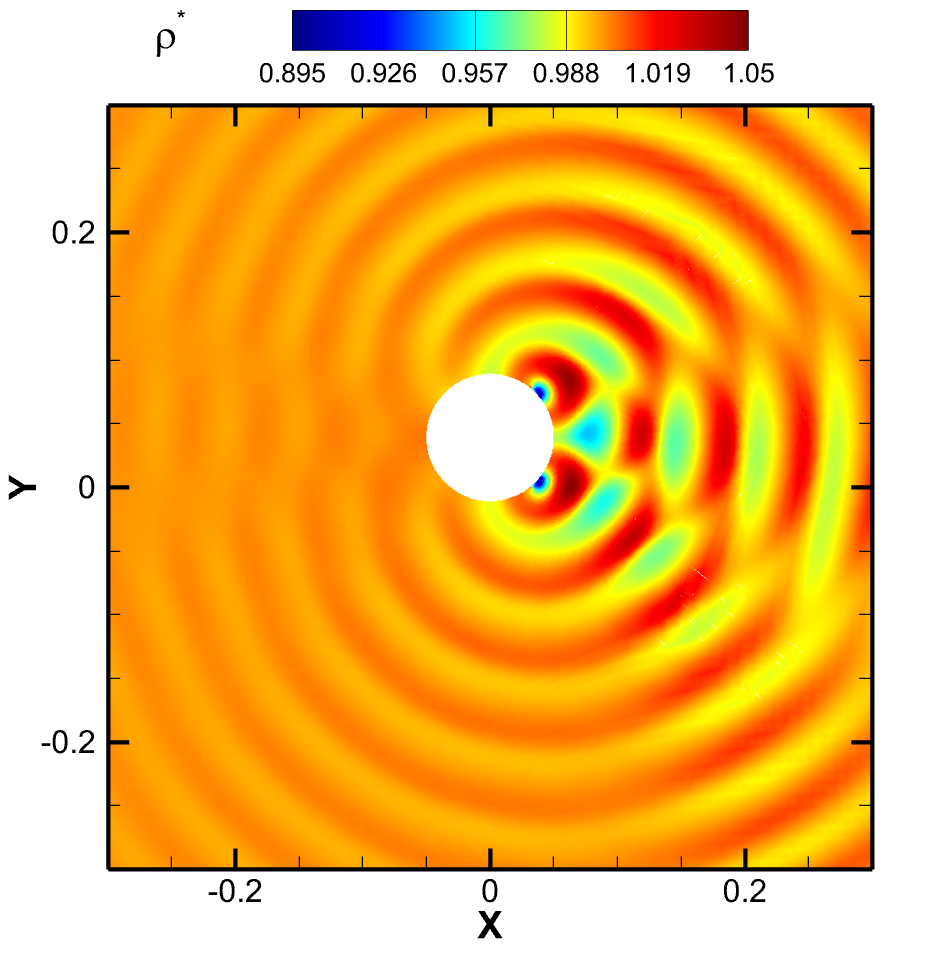} 
  \caption*{(c) $\rho^*$ contour at points $c90$}
\end{minipage}
\hfill 
\begin{minipage}[b]{0.6\textwidth}
  \centering
  \includegraphics[height=6cm]{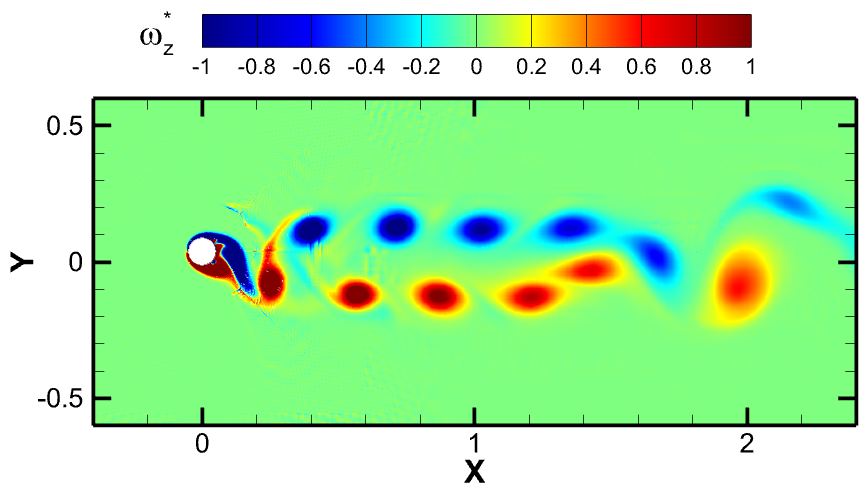} 
  \caption*{(d) $\omega_z^*$ contour at points $c90$}
\end{minipage}
\begin{minipage}[b]{0.3\textwidth}
  \centering
  \includegraphics[height=6cm]{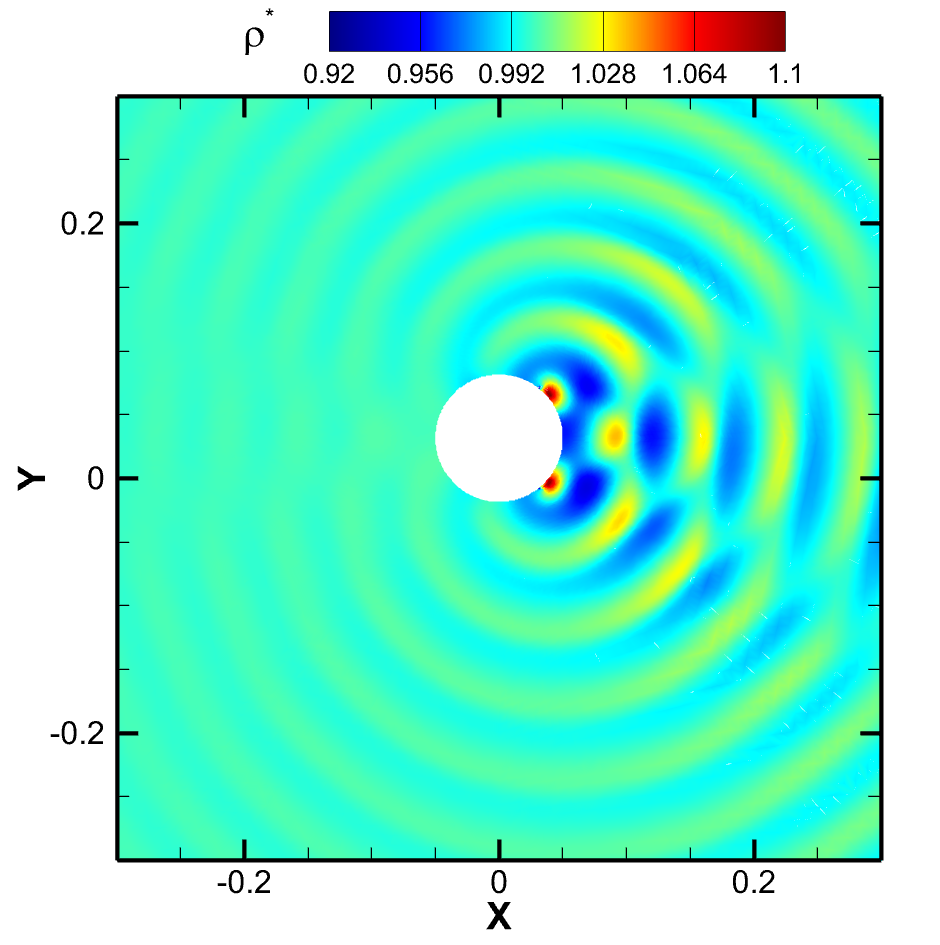} 
  \caption*{(e) $\rho^*$ contour at points $d90$}
\end{minipage}
\hfill 
\begin{minipage}[b]{0.6\textwidth}
  \centering
  \includegraphics[height=6cm]{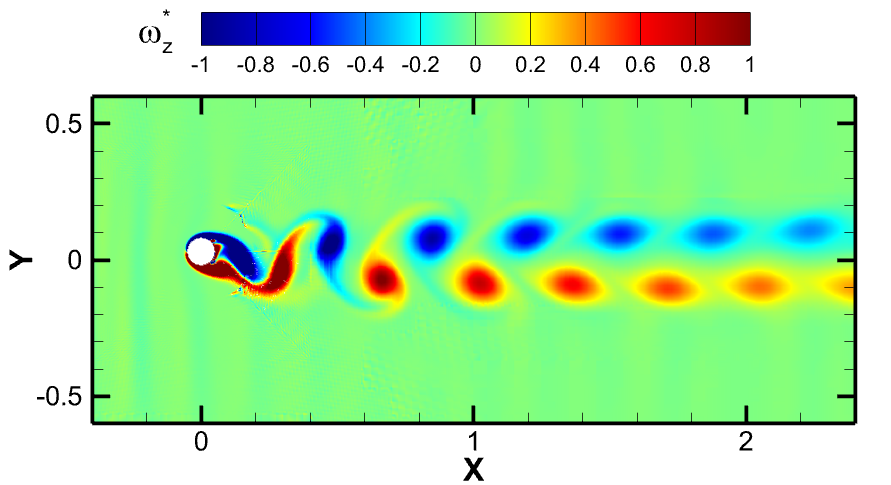} 
  \caption*{(f) $\omega_z^*$ contour at points $d90$}
\end{minipage}
  \caption{\label{fig20} $\omega_z^*$ and $\rho^*$ contour plots at points $b90$, $c90$, and $d90$}
\end{figure*}

Under the case $f^*$=90, Fig.~\ref{fig15}(d) illustrates that SJs activation 
induces a very slow attenuation of the dimensionless displacement $Y$ amplitude within 
four oscillation cycles. Beyond this initial phase, the decay rate approaches 
negligible levels, indicating amplitude stabilization at $Y_{d90}=0. 3153$. 
This behavior demonstrates further diminished efficacy in VIV suppression compared 
to lower-frequency cases. The $\omega_z^*$ and $\rho^*$ contour 
distributions at characteristic points $b90$, $c90$, and $d90$ are systematically 
presented in Fig.~\ref{fig20}. 
Analysis of the $\omega_z^*$ vorticity fields reveals that 
despite high-frequency SJs impingement, 
shed vortices along the cylinder sides exhibit no 
significant elongation or fragmentation, 
with two persistent coherent vortex streets maintained in the wake region. 
Comparative evaluation of $\rho^*$ contours demonstrates 
continued wavelength reduction in acoustic waves relative to the $f^*$=75 case. 
This phenomenon suggests predominant energy dissipation through acoustic radiation, 
where most of SJs energy propagates as acoustic waves while only 
a minimal portion is allocated to impact boundary layer for vibration mitigation.

These SJs-controlled VIV suppression cases 
demonstrate that SJs serves as an efficient and robust solution 
for vibration mitigation. Furthermore, the results reveal that SJs can achieve completely 
VIV suppression at a low actuation frequency. 
However, suppression efficiency decreases at elevated actuation frequencies due 
to the predominant conversion of SJs energy into acoustic 
waves rather than effective flow control - a finding consistent 
with the conclusions drawn by Wang et al. \cite{wang2016control} in their 2016 study. 
Finally, these cases reconfirm the computational accuracy of the 
moving mesh FSI approach developed under the RKDG AMR framework presented in this work.


\clearpage
\section{CONCLUSION}
In this paper, a moving mesh FSI approach is developed within the
RKDG AMR framework. The viscous term in the compressible NS equations 
is discretized using the high-order IPDG method. Key numerical advancements encompass the rigorous 
derivation of the Lax-Friedrichs numerical flux formulation tailored for moving meshes, an enhanced AMR-driven 
nodal correction methodology designed for curved surface geometries, and the implementation of 
a ghost-node boundary condition treatment scheme to address dynamic mesh motion.
In Couette flow simulation, the IPDG method demonstrated its high-order accuracy characteristics.
In the simulation of unsteady flow past a cylinder, the AMR technology demonstrated its capability 
to maintain high resolution while reducing computational costs.
The six VIV benchmark cases validated the accuracy of the moving mesh approach.
Furthermore, the proposed moving mesh FSI approach was applied to investigate the VIV suppression
of a single cylinder under active flow control.
The results demonstrated that SJs can effectively suppress VIV at a low actuation frequency,
while higher actuation frequencies lead to reduced suppression efficiency due to the energy of the SJs
is more in the form of acoustic wave.The following conclusions are obtained:

1. The RKDG AMR framework demonstrates superior numerical stability and high-order accuracy characteristics, where the AMR technology effectively optimizes computational resource allocation while preserving solution fidelity. 
\\

2. The IPDG algorithm exhibits significant computational advantages in hydrodynamic simulations through its inherent high-order spatial resolution capabilities. 
\\

3. The moving mesh FSI approach developed within the RKDG AMR framework has exceptional performance in VIV simulations. 
\\

4. SJs is an efficient and robust solution for vibration mitigation, SJs can achieve completely 
VIV suppression at a low actuation frequency, while higher actuation 
frequencies reduce suppression efficiency due 
to the energy of the SJs is more in the form of acoustic wave. 
\section*{ACKNOWLEDGMENTS}
**************
\section*{DATA AVAILABILITY}
The data that support the findings of this study are available 
within the article. 
\section*{REFERENCES}
\nocite{*}
\bibliography{aipsamp}

\end{document}